\documentclass{emulateapj}
\def\be{\begin{equation}}
\def\ee{\end{equation}}
\def\bea{\begin{eqnarray}}
\def\eea{\end{eqnarray}}

\begin{document}
\title{Physical processes shaping GRB
X-ray afterglow lightcurves: theoretical implications
from the {\em Swift} XRT observations} 

\author{Bing Zhang$^{1}$, Y. Z. Fan$^{1,2,3}$, Jaroslaw Dyks$^{1,4}$,
Shiho Kobayashi$^{5,6,7}$, Peter M\'{e}sz\'{a}ros$^{5,6}$, \\
David N. Burrows$^{5}$, John A. Nousek$^{5}$, 
and Neil Gehrels$^{8}$} 
\affil{$^1$ Dept. of Physics, University of Nevada, Las Vegas, NV
89154.\\
$^2$ Purple Mountain Observatory, Chinese Academy of
Science, Nanjing 210008, China.\\
$^3$ National Astronomical Observatories, Chinese Academy of
Sciences, Beijing, 100012, China.\\
$^4$ Nicolaus Copenicus Astronomical Center, Toru\'n, Poland. \\
$^5$ Dept. of Astronomy and Astrophysics, Pennsylvania State
University, 525 Davey Laboratory, University Park, PA 16802.\\ 
$^6$ Dept. of Physics, Pennsylvania State
University, 104 Davey Laboratory, University Park, PA 16802.\\
$^7$ Astrophysics Research Institute, Liverpool John Moores
University, Twelve Quays House, Birkenhead, CH41 1LD, UK. \\
$^8$ NASA/Goddard Space Flight Center, Greenbelt, MD 20771.
} 

\begin{abstract}
With the successful launch of the {\it Swift} Gamma-ray Burst
Explorer, a rich trove of early X-ray afterglow data has been
collected by its on-board X-Ray Telescope (XRT). Some interesting
features are emerging, including a distinct rapidly decaying component
preceding the conventional afterglow component in many
sources, a shallow decay component before the more ``normal'' decay
component observed in a good fraction of GRBs (e.g. GRB 050128, GRB
050315, GRB 050319, and GRB 050401), and X-ray flares in nearly
half of the afterglows (e.g. GRB 050406, GRB 050502B, GRB 050607, and GRB
050724). These interesting early afterglow signatures reveal valuable
and unprecedented information about GRBs, including the prompt
emission - afterglow transition, GRB emission site,
central engine activity, forward-reverse shock physics, and 
the GRB immediate environment. In this paper, we systematically
analyze the possible physical processes that shape the properties of
the early X-ray afterglow lightcurves, and use the data to constrain
various models. We suggest that the steep decay component is
consistent with the tail emission of the prompt gamma-ray bursts
and/or of the X-ray flares. This provides strong evidence that 
the prompt emission and afterglow emission are likely two 
distinct components, supporting the internal origin of the
GRB prompt emission. The shallow decay segment observed in a group of
GRBs suggests that very likely the forward shock keeps being
refreshed for some time. This might be caused either by a long-lived
central engine, or by a wide distribution of the shell Lorentz
factors, or else possibly by the deceleration of a Poynting flux
dominated flow. X-ray flares suggest that the GRB central engine is very
likely still active after the prompt gamma-ray emission is over, but
with a reduced activity at later times. In some cases, the central
engine activity even extends to days after the burst
triggers. Analyses of early X-ray afterglow data reveal that GRBs 
are indeed highly relativistic events, and that early
afterglow data of many bursts, starting from the beginning of the XRT
observations, are consistent with the afterglow emission from an
interstellar medium (ISM) environment. 
\end{abstract}



\section{Introduction}
With the successful launch of the {\em Swift} Gamma-ray Burst
Explorer, an era of systematic, multi-wavelength observations of
GRB early afterglows has been ushered in. Very early optical/IR
detections have been made with ground-based telescopes (Akerlof et
al. 1999; Fox et al. 2003; Li et al. 2003; Blake et al. 2005;
Vestrand et al. 2005) before and during the initial operation of
{\em Swift}. In the X-ray band, some evidence of the early afterglows
has been collected earlier (e.g. Piro et al. 1998; Giblin et al. 1998;
Burenin et al. 1999; Piro et al. 2005). However, it is the operation
of the {\em Swift} X-Ray Telescope (XRT) that offers the possibility
to unveil the final gap between the prompt emission and the late
afterglow stage.  

There has been widespread expectation that the early X-ray
observations could answer a series of core questions in GRB
studies. What is the 
connection between the GRB prompt emission and the afterglow? Are
prompt emission and afterglow both from the external shock
(M\'esz\'aros \& Rees 1993; Dermer \& Mitman 1999) or do they come from
different locations [i.e. prompt emission from the internal shocks
(Rees \& M\'esz\'aros 1994; Paczynski \& Xu 1994), while the
afterglow comes from the external shock (M\'esz\'aros \& Rees 1997a;
Sari et al. 1998)]? Does the central engine become
dormant after the burst is over? What is the immediate environment of
the burst, an ISM or a wind? Are there density clumps surrounding the
GRB progenitor? What is the role of
the reverse shock? What is the initial Lorentz factor of the fireball?

All these questions can be at least partially answered with the early
X-ray afterglow data, sometimes in combination with the prompt
gamma-ray data and the early optical/IR afterglow data. 
Although early afterglow lightcurves have been extensively modeled in
the optical band (mainly driven by the observations and by the
theoretical argument that the reverse shock emission component plays
an important role in the optical band, e.g. M\'esz\'aros \& Rees
1997a; Sari \& 
Piran 1999; M\'esz\'aros \& Rees 1999; Kobayashi 2000; Kobayashi \&
Zhang 2003a,b; Zhang et al. 2003; Wei 2003; Wu et al. 2003; Li et
al. 2003b; Fan et al. 2004a; Zhang \& Kobayashi 2005; Fan et
al. 2005a,b; Nakar \& Piran 2004; McMahon et al. 2004), possible early
X-ray afterglow signatures have been only sparsely studied (e.g. Kumar
\& Panaitescu 2000a; Kobayashi et al. 2005a; Fan \& Wei 2005).

In its first six months of operations, the {\em Swift} X-Ray Telescope
(XRT) has already accumulated a rich collection of early afterglow
features in the X-ray band. The XRT is a sensitive broad-band
(0.2-10 keV) imager, which can be promptly slewed to GRB targets
triggered by the Burst Alert Telescope (BAT) within 1-2 minutes 
(Burrows et al. 2005a). It is therefore an ideal instrument
to study the transition between the GRB prompt emission and the very 
early X-ray afterglow.
The following features are all detected by XRT in a good sample of
bursts (see, e.g. Chincarini et al. 2005; Nousek et al. 2005 for a
collection of data), reflecting some common underlying physics of GRBs.

1. In most cases (e.g. GRB 050126 and GRB 050219A), a steep decay is
observed up to several hundred seconds after the burst trigger, which
is followed by a more conventional shallower decay (Tagliaferri et
al. 2005; Goad et al. 2005). This conclusion is drawn by choosing
the GRB trigger time as the zero time point ($t_0$). At later times 
(e.g. $(t-t_0) \gg T_{90}$, where $T_{90}$ is the duration of the GRB), 
the afterglow decay slope ${\rm d} \ln F_\nu/ {\rm d} \ln (t-t_0)$ is
essentially independent on the adopted $t_0$. However, at early times
(e.g. $(t-t_0)$ not much larger than $T_{90}$, the measured
decay slope could be very sensitive to the assumed $t_0$. 
Tagliaferri et al. (2005) explored the $t_0$ effect and concluded that
the two distinct lightcurve segments are likely intrinsic rather than
due to a poor choice of $t_0$. Furthermore, in some cases, the steep decay
segment also has a different spectral index (e.g. for GRB 050319,
Cusumano et al. 2005). Usually it also connects to the spectral
extrapolation of the BAT prompt emission lightcurve smoothly
(Barthelmy et al. 2005a).  All these facts indicate that the steep decay
component is a distinct new component that may be unrelated to the
conventional afterglow component.

2. In a good fraction of GRBs (e.g. GRB 050128, Campana et al. 2005;
GRB 050315, Vaughan et al. 2005; GRB 050319, Cusumano et al. 2005; and
GRB 050401, de Pasquale et al. 2005), the early X-ray afterglow is
characterized by a shallow to ``normal'' transition. During the
transition the spectral index is not changed. The decay slope after the
break (e.g. $\sim -1.2$) is consistent with the standard ISM afterglow
model, while the decay slope before the break is usually much
shallower (e.g. $\sim -0.5$).

3. In some cases (e.g. GRB 050315, Vaughan et al. 2005), a further
steepening is detected after the ``normal'' decay phase, which is
consistent with a jet break.

4. Bright X-ray flares have been detected in the early X-ray
lightcurves of nearly one half of the burst population 
(e.g. GRB 050406, GRB 050202B, Burrows et al. 2005; Romano et
al. 2005; Falcone et al. 2005). In particular, the X-ray afterglow of
the short-hard burst GRB 050724 also shows at least three flares 
(Barthelmy et al. 2005b). The flares typically happen hundreds of
seconds after the trigger or earlier, but in some cases they
occur around a day (e.g. GRB 050502B, Falcone et al. 2005; GRB 050724,
Barthelmy et al. 2005b). The amplitudes of the flares are usually
larger than the underlying afterglow component by a factor of several 
(e.g. a factor of 6 in GRB 050406, Burrows et al. 2005; Romano et
al. 2005), but can be much larger (e.g. $\sim 500$ in the case of GRB
050202B, Burrows et al. 2005; Falcone et al. 2005). A similar
feature was evident for GRB 011121 detected by BeppoSAX (Piro et
al. 2005). 

\begin{figure}
\includegraphics[angle=-90,scale=0.35]{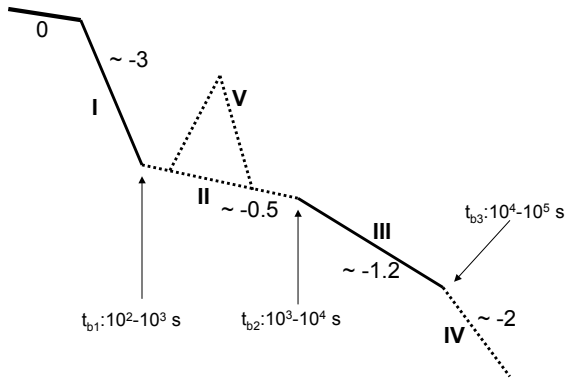}
\caption{A synthetic cartoon X-ray lightcurve based on the
observational data from the {\em Swift} XRT. The phase ``0'' denotes
the prompt emission. Four power law lightcurve segments together with
a flaring component are identified in the afterglow phase.  
The segments I and III are most common, and they are marked with solid
lines. Other three components are only observed in a fraction of
bursts, so they are marked as dashed lines. Typical temporal
indices in the four segments are indicated in the figure. The spectral
indices remain unchanged for segments II, III and IV, with a typical 
value of $\beta_X \sim 1$ ($F_X \propto \nu^{-\beta_X}$). The segment
I sometimes has a softer spectrum (e.g. $\beta_X \sim 1.5$),
but in some other cases it has a similar spectral index as the other
three segments. The flares (segment V) have similar spectra as the
segment I, and time evolution of the spectral index during the flares
has been observed in some bursts (e.g. GRB 050502B).} 
\label{XRTlc}
\end{figure}

In summarizing the current X-ray afterglow data, one can tentatively
draw a synthetic cartoon lightcurve in the X-ray band, which consists
of 5 components (see Figure \ref{XRTlc}): I. an initial steep decay
(with a typical slope $\sim -3$ or steeper); II. a shallower-than-normal
decay (with a typical slope $\sim -0.5$); III. a normal decay (with a
typical slope $\sim -1.2$); IV. a late steeper decay (with a typical
slope $\sim -2$); and V. one or more X-ray flares. We note that
Nousek et al (2005) also arrived at the similar schematic diagram that
includes the segments I, II and III in our cartoon picture (see their 
Fig.3). Limited by the 
quality of the data, the current analyses indicate that the spectral 
indices remain unchanged in the segments II, III and IV, with a
typical value of $\beta_X \sim 1$ ($F_X \propto
\nu^{-\beta_X}$, Nousek et al. 2005). In some bursts, the segments I
and II have different spectral indices (e.g. GRB 050319, Cusumano
et al. 2005). In some cases, a time-evolution of the spectral index
has been detected (e.g. in the giant flare of GRB 050502B, Falcone et
al. 2005). In Fig.1, we have indicated the typical temporal index for
each segment. Throughout the paper
the transition times between the adjacent segments for 
the four lightcurve segments are denoted as $t_{b1}$, $t_{b2}$, and
$t_{b3}$, respectively.
In this paper, we systematically study the physical processes
that shape an early X-ray lightcurve, and discuss possible
theoretical interpretations to the above phenomena. In
\S{\ref{sec:Curvature}}, we discuss the GRB tail emission
arising from high angular latitude relative to the viewing direction,
which takes longer to reach the observer due to the extra
distance it travels, as a conical (or spherical) shell suddenly stops
shining. This is the so-called ``curvature effect" (Fenimore et al. 1996;
Kumar \& Panaitescu 2000a; Dermer 2004; Dyks et al. 2005).  In
\S{\ref{sec:FS}}, we review the main 
emission properties from the external forward shock region in the
X-ray band, summarizing the temporal and spectral indices expected in
the X-ray band for both the ISM and the wind models. Furthermore, we
also discuss the case of a continuously refreshed shock as well as its
three possible physical mechanisms. Several case studies are
investigated to reveal an intriguing refreshed shock phase commonly
present in many 
bursts. In \S{\ref{sec:RS}}, we briefly discuss whether and how the
reverse shock emission would influence the X-ray band emission. 
In \S{\ref{sec:flare}}, we explore 
various mechanisms that might give rise to the X-ray flares observed
in many bursts (e.g. GRB 050406 and GRB 050202b), and conclude that
the phenomenon is best interpreted as due to a late central engine
activity. Our conclusions are summarized in \S{\ref{sec:conclusions}}.


\section{GRB tail emission and the curvature effect}
\label{sec:Curvature}  

\subsection{GRB tail emission}

The temporal bridge between the GRB prompt emission and the afterglow
emission is essential for revealing whether the prompt emission and
the afterglow originate from the same component. The earliest GRB
relativistic shock model invoked the external shock as the site for
prompt $\gamma$-ray emission (Rees \& M\'esz\'aros 1992; M\'esz\'aros
\& Rees 1993). The rapid variability observed in many GRBs is in
great contrast with the intuitive expectations in the external shock
model, which generally predicts a smooth burst temporal profile, and
it has been argued that the radiative efficiency is too low for the
model so that a much larger energy budget is required (Sari \& Piran
1997). Dermer \& Mitman (1999, 2003) argued that if
the GRB ambient medium is sufficiently clumpy, an external shock GRB
model could reproduce the observed variability with a high energy
efficiency. Within such a picture, the prompt emission and the
afterglow originate from the same component (i.e. the external
shock), and it is expected that the two emission components are likely
smoothly connected in the early phase.

On the other hand, it is now commonly believed that GRB prompt
emission originates from some ``internal processes'', i.e. the
$\gamma$-rays are emitted before the fireball is decelerated by the
ambient medium.  The most widely discussed
model is the internal shock model (Rees \& M\'esz\'aros 1994;
Paczynski \& Xu 1994; Kobayashi et al. 1997; Daigne \& Mochkovitch
1998; Fan et al. 2004b). Alternatively, the internal emission may be
caused by dissipation of strong magnetic fields (e.g. Drenkhahn \&
Spruit 2002), or Comptonization of the photospheric emission (Rees \&
M\'esz\'aros 2005). Within such scenarios, there exist two distinct
temporal episodes dominated by the prompt emission and the afterglow,
respectively, since the latter is emitted at a much larger distance
from the central engine when the fireball is decelerated. Generally
one should expect a flux contrast between these two episodes. 

Before the {\em Swift} era, no solid observation was available to
finally differentiate both scenarios, and evidence in favor of each
scenario had been collected (see e.g. Zhang \& M\'esz\'aros 2004 for a
review). It is one of the major tasks of {\em Swift} to pin
down the emission site of the GRB prompt emission.

If the prompt emission and the afterglow arise from different emission
sites, as is expected in the internal shock (or similar) scenario, and
if the prompt emission flux level is much higher than 
the afterglow emission flux level, one expects to see a steeply decaying
lightcurve during the transition from the prompt emission phase to the 
afterglow phase. Such a steep decay is due to the so-called
``curvature effect'' (e.g. Kumar \& Panaitescu 2000a; Dermer 2004;
Dyks et al. 2005; Panaitescu et al. 2005).  In principle,
such an effect also applies to the tail emission of the X-ray flares 
(Burrows et al. 2005).  Hereafter we generally define such an emission
component as ``GRB tail emission''.

\subsection{Curvature effect}

\subsubsection{The simplest case}

For a conical jet with an opening angle $\theta_{\rm j}$, emission
from the same radius $R_{cr}$ but from different viewing latitudes
$\theta$ ($\theta<\theta_{\rm j}$) would reach the observer at
different times. Even if the emission is turned off instantaneously,
due to the propagation effect the observer would receive the emitted
photons at the angle $\theta$ at $t = (1+z) (R_{cr}/c)
(\theta^2/2)$. Such a tail emission thus lasts for a duration of 
\be
t_{tail} = (1+z) (R_{cr}/c) (\theta_{\rm j}^2/2) \simeq 330~{\rm s}
\left(\frac{R_{cr} \theta_j^2}{10^{13}}\right) \left(\frac{1+z}{2}
\right). 
\ee 
if the line of sight is not too
close to the jet edge.  

We consider the simplest case of a jet moving with a constant bulk
Lorentz factor $\Gamma$ (or a constant velocity $v$). The electrons
are shock-heated up to a radius $R_{cr}$, beyond which no fresh
shocked electrons are injected, and the already heated electrons cool
rapidly. The comoving emission frequency $\nu'$ is boosted to $\nu =
{\cal D} \nu'$ in the observer's frame, where ${\cal D}=[\Gamma
(1-v\cos\theta/c)]^{-1}$ is the Doppler factor, which is ${\cal D}
\sim 2 \Gamma$ for $\theta \ll 1/\Gamma$, and ${\cal D} \sim 2/(\Gamma
\theta^2)$ for $\theta \gg 1/\Gamma$. Since $t \propto \theta^2$, one
gets ${\cal D} \propto t^{-1}$ for $\theta \gg 1/\Gamma$. 

The observed flux $F_\nu$ is related to the comoving surface
brightness $L'_{\nu'}$ by 
\be
F_\nu \propto L'_{\nu'} {\cal D}^2 \propto (\nu')^{-\beta} {\cal D}^2
\propto \nu^{-\beta} {\cal D}^{2+\beta} \propto \nu^{-\beta}
t^{-2-\beta}~, 
\ee
where $\beta$ is the observed spectral index around the observed
frequency $\nu$, and the last proportionality is valid for $1/\Gamma
\ll \theta < \theta_j$. With the standard convention $F_\nu \propto 
\nu^{-\beta} t^{-\alpha}$, one has the well-known result for the
curvature effect (e.g. Kumar \& Panaitescu 2000a; Dermer 2004; Fan \&
Wel 2005; Dyks et al. 2005; Panaitescu et al. 2005)
\be
\alpha=2+\beta~.
\ee

\bigskip
\subsubsection{Emission from a decelerating fireball}
\begin{figure}
\epsscale{1.0}
\plotone{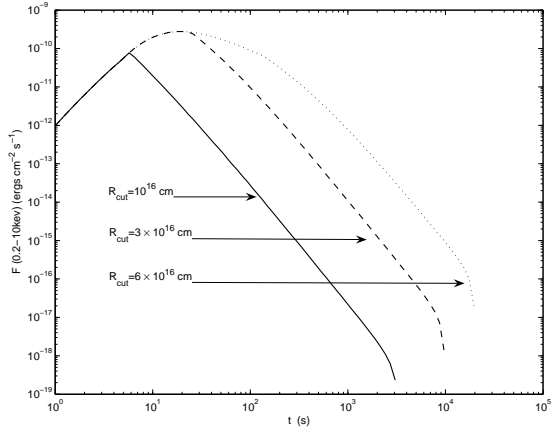}
\caption{The curvature effect for a decelerating ejecta. In the
calculation, the radiation from the fluid elements on the equal
arrival time surface for $R>R_{cut}$ are cut out, and the
radiation from all other fluid elements is integrated. Here
$R_{cut}=10^{16}, 3\times 10^{16}, 6\times 10^{16}$ cm are the radii
at which the radiation is assumed to terminate abruptly. Notice
the insensitivity of the decay slope ($\alpha=2+\beta$) on the
cut-off radius. The following parameters are adopted: the isotropic
kinetic energy $E_{iso}=10^{52}$ergs, the initial Lorentz factor
$\Gamma_0=240$, the jet half-opening angle $\theta_{\rm j}=0.1$,  
the ambient density $n=1{\rm cm^{-3}}$, the redshift $z=1$ (the
luminosity distance $D_L=2.2 \times 10^{28}$cm), the electron spectral
index $p=2.3$, and the electron and magnetic equipartition parameters
$\epsilon_e=0.1$ and $\epsilon_B=0.01$, respectively.}. 
\label{Fig:Cur-ISM}
\end{figure}

In reality, the Lorentz factor of the shell could be decreasing
right before the sudden cessation of the emission. This is valid for
the external shock case in the deceleration phase, or even in the
internal shock case. We perform numerical calculations to investigate
such an 
effect\footnote{In this paper, we use two codes to perform numerical 
calculations. The first code was developed by J. Dyks (Dyks et
al. 2005). The code can deal with the afterglow emission of an outflow
with an arbitrary axisymmetric structure and an arbitrary observer's
viewing direction. The dynamics of the radial outflow 
is modeled similar to Granot \& Kumar (2003). Only synchrotron
radiation is taken into account at the moment and synchrotron
self-absorption (which is irrelevant for X-ray emission) is
ignored. The synchrotron 
emissivity in the comoving frame is calculated by integration of the
synchrotron spectrum over the electron energy distribution. The latter
is calculated by solving the continuity equation with the power-law
source function $Q=K\gamma^{-p}$, normalized by a local injection rate
(Moderski et al. 2000). The electrons cool down through synchrotron
radiation and 
adiabatic expansion. All kinematic effects which affect the observed
flux (e.g. Doppler boost, propagation time effects) have been taken
into account rigorously, following Salmonson (2003). The second code
was developed by Y. Z. Fan. This is an afterglow code developed from
the dynamical model of Huang et al. (2000), and has been used in
several previous studies (e.g. Fan et al. 2004a; Fan et
al. 2005a). The latest addition is to also include the kinetic
evolution of the electron distribution (Moderski et al. 2000). We have
used both codes in various calculations in this paper, and the results
are consistent with each other. The figures presented in this paper
are all generated from the second code. }.
The curvature effect for the sudden switching-off of radiation in
a decelerating outflow is presented Figure \ref{Fig:Cur-ISM}, in which
we plot the X-ray lightcurve from an expanding jet blastwave with
$E_{iso}=10^{52}$ergs, $\Gamma_0=240$, $\theta_j = 0.1$, $n=1{\rm
cm^{-3}}$, and $z=1$. We manually turn off the radiation at radii
($R_{cut}$) of $10^{16}$ cm, $3\times 10^{16}$ cm, and $6\times
10^{16}$ cm, respectively, and investigate the subsequent curvature
effect. The radiation from fluid elements at $R<R_{cut}$ has been 
rigorously integrated. The result indicates that the $\alpha =
2+\beta$ conclusion is essentially unchanged.
\subsubsection{Jet structure effects}
Another interesting issue is the jet structure. In principle, GRB jets
could be structured (Zhang \& M\'esz\'aros 2002a; Rossi et al. 2002;
Zhang et al. 2004a; Kumar \& Granot 2003). Since the curvature effect
allows one to see the high-latitude emission directly, an interesting
question is whether the decay slope associated with the curvature
effect depends on the unknown jet structure. We have investigated this
effect with the first code, and find that for a relativistic outflow
the temporal slope of the curvature effect is largely insensitive to
the jet structure as long as the viewing angle is not far off the
bright beam. The main reason is
that the decrease of flux because of the curvature effect occurs on
a much shorter timescale than that for the jet structure to take
effect. For a spectral index $-\beta$ ($f_\nu \propto \nu^{-\beta}$), 
the flux decreases by $m$ orders of magnitude after a time of $t_{\rm
crv}=10^{m/(2+\beta)}t_{cr}$, where $t_{cr}$ is the observer time at
which the curvature effect began. For a typical $\beta \sim 1$,
the flux drops by one order of magnitude after a short time $t_{\rm
crv}\sim 2t_{cr}$. A drop of three orders of magnitude occurs in no
more than a decade in time. On the other hand, the observer can
perceive the switch-off of emissivity at an angle $\theta$ measured
from the line of sight at a time $t_{\rm \theta} \simeq [1 +
(\theta\Gamma)^2]t_{cr}$. One can see that the structure of the outflow
must have a typical angular scale smaller than $3/\Gamma$ in order
to affect the observed flux before $10t_{cr}$. For $\Gamma > 10^2$, the
parameters of the outflow would have to vary strongly on a scale
smaller than one degree. Nonetheless, the effect of the jet struture
would start to play a noticeable role if the line of sight is outside
the bright beam. Detailed calculations are presented elsewhere (Dyks
et al. 2005).

\subsubsection{Factors leading to deviations from the $\alpha=2+\beta$
relation} 

Almost all the {\em Swift} XRT early afterglow lightcurves are
categorized by a steep decay component followed by a more ``normal'' 
decaying afterglow lightcurve (Tagliaferri et al. 2005; Nousek et
al. 2005) - see Segment 
1 in Fig.\ref{XRTlc}. In most of these cases, the measured $\alpha$
and $\beta$ values in this rapidly decaying component are close to the
$\alpha=2+\beta$ relation, but in most cases, they do not match
completely. This does not invalidate the curvature effect
interpretation, however, since in principle the following factors
would lead to deviations from the simple $\alpha=2+\beta$ law. 

\begin{figure}
\epsscale{1.0}
\plotone{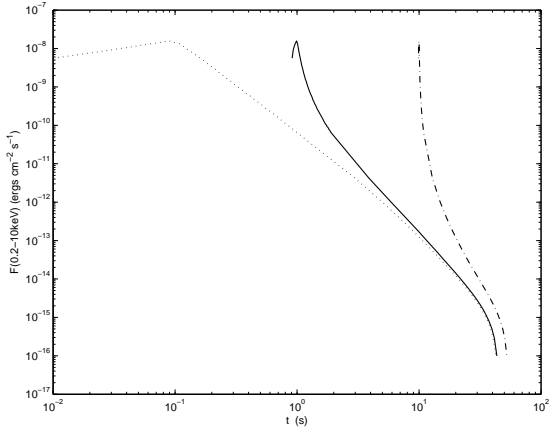}
\caption{The effect of $t_0$ on the lightcurves. The same internal shock
$\gamma$-ray pulse is calculated, but is assigned to three different
ejection times $t_{ej}$. The dotted, solid, and dash-dotted lines are
for $t_{\rm ej}=0.0,~1.0,~10.0$s, respectively. The following parameters
are adopted to calculate the internal shock pulses: the pulse
luminosity $L_{\rm pulse}=10^{51}{\rm ergs~s^{-1}}$, the variability
time scale $\delta t=0.1$s, $\theta_{\rm j}=0.2$rad, $\epsilon_{\rm
e}=0.5$, $\epsilon_{\rm B}=0.1$, $p=2.5$, and $z=1$.} 
\label{Fig:Internal}
\end{figure}

1. The time zero point ($t_0$) effect. In GRB studies, the afterglow 
lightcurves are plotted in the log-log scale, with $t_0=0$ defined as
the trigger time of the burst. When discussing the late afterglows,
shifting $t_0$ by the order of the burst duration $T_{90}$ does not
make much difference. When discussing the early afterglow and its
connection to the prompt emission, however, the decay power law
index (${\rm d} \ln F_\nu / {\rm d} \ln t$) is very sensitive to the
$t_0$ value one chooses. Correctly 
choosing $t_0$ is therefore essential to derive the correct temporal
decay index $\alpha$ (see e.g. Kobayashi et al. 2005b
for a detailed discussion about the $t_0$ issue). In the case of the
internal shock 
model, the case is straightforward. The observed pulses essentially
track the behavior of the central engine (Kobayashi et al. 1997). 
Each pulse marks another re-start of the central engine, so that
$t_0$ should be re-defined for each pulse when the curvature effect of
that pulse is considered. Keeping the same $t_0$ as the beginning of the
first pulse (i.e. the GRB trigger) would inevitably lead to a false,
very steep power law decay for later pulses. Figure \ref{Fig:Internal}
gives an example to show this point. When a series of shells are
successively ejected, each pulse will be followed by its tail emission
due to the curvature effect, but most of these tails are buried under the
main emission of the next pulse. The observed curvature effect is only
the tail emission of the very last pulse. As a result, properly
shifting $t_0$ is essential to interpret the steep decay component
observed in XRT bursts.
2. The superposition effect. The observed steep-to-shallow transition
in the early phase of XRT bursts (Tagliaferri et al. 2005) suggests
that by the end of the tail emission, the fireball is already
decelerated, and the forward shock emission also contributes to the
X-ray band. As a result, the observed steep decay should also include
the contribution from the forward shock. Assuming the later has a
temporal decay index $-w$, the X-ray flux at the early phase should
read 
\begin{equation}
F_{\nu}(t)=A\left(\frac{t-t_{0,i}}{t_{0,i}}\right)^{-(2+{\beta})}
+B\left(\frac{t-t_{0,e}}{t_{0,e}}\right)^{-w}, 
\label{Eq:Fnu_obs}
\end{equation}
where $A$ and $B$ are constants, and $t_{0,i}$, $t_{0,e}$ are the
time zero points for the steep decay component (presumably of the
internal origin) and for the shallow decay component (presumably of
the external origin), respectively. 
In the intermediate regime between the two power-law segments, both
components are important, and the observed $\alpha$ should be
shallower than $2+\beta$ during the steep decaying phase. This effect
flattens the decay instead of steepening it. 

3. If with the above two adjustments, the observed $\alpha$ is
still steeper than $2+\beta$, one can draw the conclusion that the
solid angle of the emitting region is comparable to or smaller than
$1/\Gamma$. This would correspond to a patchy shell (Kumar \& Piran
2000a) or a mini-jet (Yamazaki et al. 2004). A caveat on such an
interpretation is that the probability for the line of sight sitting
right on top of such a very narrow patch/mini-jet is very small.
As a result, this model can not interpret an effect that seems to be a
general property of X-ray afterglows.

4. If with the first two adjustments, the observed $\alpha$ is
flatter than $2+\beta$ but is still much steeper than that expected
from a forward shock model, there could be two possibilities. One is
that the emission is still from the internal dissipation of energy but
the emission in the observational band does not cease abruptly. This
is relevant when the observational band is below the cooling
frequency. The adiabatic cooling therefore gives a decay slope of
$\sim (1+3\beta/2)$ rather than $\sim (2+\beta)$ (e.g. Sari \& Piran
1999; Zhang et al. 2003). The second possibility is that one is
looking at a structured jet (Zhang \& M\'esz\'aros 2002a; Rossi et
al. 2002), with the line of sight significantly off-axis. The
curvature effect in such a configuration typically gives a flatter
decay slope than $2+\beta$ (Dyks et al. 2005). This is particularly
relevant for X-ray rich GRBs or X-ray flashes for which a large
viewing angle is usually expected (Zhang et al. 2004; Yamazaki et
al. 2004). Further analyses of XRT data suggest that at least in some
GRBs, the decay slope is shallower than $2+\beta$ (O'Brien et
al. 2005). The above two possibilities are in particular relevant for
these bursts.

We suggest that most of the rapid-decay lightcurves observed by the
{\em Swift} XRT may be interpreted as GRB (or X-ray flare) tail
emission through the curvature effect, with the first two adjustments
discussed above. In order to test this hypothesis, after the
submission of this paper we have performed more detailed data analyses
on a large sample of XRT bursts (Liang et al. 2005). By assuming that
the decay slope should be $2+\beta$, we search the appropriate $t_0$
that allows such an assumption to be satisfied. It is found that $t_0$
is usually at the beginning of the last pulse (for the steep decay
following the prompt emission) or at the beginning of the X-ray flare
(for steep decay following flares). This fact strongly suggests that
the curvature effect is likely to be the correct interpretation, and
at the same time lends strong support to the internal origin of the
prompt emission and X-ray flares (see \S\S2.3 \& 5.7 for the arguments
in favor of the internal models for both components).

Another potential test of the curvature effect is to search for a
correlation between the spectral peak energy ($E_{pk}$) and the flux
at the peak ($F_{pk}$) (Dermer 2004). This requires a well measured
$E_{pk}$ in the XRT band. In most cases, the XRT spectrum is
consistent with a single power law. More detailed analyses on future
bright X-ray flares are desirable to perform such a test.

\subsection{Theoretical implications}

The current {\em Swift} XRT observations of the early rapid-to-shallow
decay transition of the X-ray lightcurves (Tagliaferri et al. 2005),
when interpreted as the curvature effect, have profound
implications for the understanding of the GRB phenomenon. 

1. It gives a direct observational proof that the GRB prompt emission
very likely comes from a different site than the afterglow
emission. This suggests that the emission either comes from the
internal shocks due to the collisions among many engine-ejected shells
(Rees \& M\'esz\'aros 1994; Paczynski \& Xu 1994), or is due to
magnetic or other dissipation processes at a radius smaller than the
fireball deceleration radius (e.g. Drenkhahn \& Spruit 2002; Rees \&
M\'esz\'aros 2005). In both scenarios, the energy dissipation region
is well inside the region where the deceleration of the whole fireball
occurs. 

2. An interesting fact is that in most cases, after the
prompt emission, the X-ray emission level (that spectrally
extrapolated from the BAT data) drops by several orders of
magnitude (through the curvature effect, in our interpretation) before
``landing'' on the  
afterglow emission level. One could roughly estimate the expected
``drop-off''. The flux level in the XRT band during the prompt phase
could be roughly estimated as $F_{\nu,X}^{prompt} \propto 
(E_{\gamma,iso}/T_{90}) (E_{\rm XRT}/E_p)^{\hat\alpha+2}$, where
$E_{\gamma,iso}$ is the isotropic energy of the emitted gamma-rays,
$T_{90}$ is the duration of the burst, $E_{\rm XRT} \sim 5$ keV is the
typical energy in the XRT band, $E_p \sim 100$ keV is the typical
peak energy in the GRB spectrum, and $\hat\alpha \sim -1$ is the
low-energy spectral index for a Band-spectrum (Band et al. 1993).
Assuming that the X-ray band for the afterglow emission is above both
the typical synchrotron 
frequency $\nu_m$ and the cooling frequency $\nu_c$ (which 
is usually the case for the ISM model, see eqs.[\ref{tm}],[\ref{tc}]),
the X-ray afterglow flux level can be estimated as (e.g. Freedman \&
Waxman 2001) $F_{\nu,X}^{ag} \propto \epsilon_e E_{iso}/t$, where
$E_{iso}$ is the isotropic energy of the afterglow kinetic energy, and
$\epsilon_e$ is the electron equipartition parameter in the
shock\footnote{When the synchrotron self-Compton process dominates the
cooling, the discussion could be more complicated.}.
The flux contrast can be estimated as
\be
\frac{F_{\nu,X}^{prompt}} {F_{\nu,X}^{ag}} \sim \left(\frac{E_{\gamma,iso}}
{E_{iso}}\right) \left(\frac{t}{T_{90}}\right) \left[\frac{(E_{\rm
XRT}/E_p)^{\hat\alpha+2}}{\epsilon_e}\right]~.
\ee
For typical parameters, one has $(E_{\rm XRT}/E_p)^{\hat\alpha+2} \sim
0.05$ and $\epsilon_e \sim 0.1$, so that the term in the bracket
$\lesssim 1$. Although $t>T_{90}$ would generally suggest that
$F_{\nu,X}^{prompt}$ should be higher than $F_{\nu,X}^{ag}$, the
large contrast between the two components observed in many bursts
is usually not accounted for unless $E_{\gamma,iso}$ is (much) larger
than $E_{iso}$. This refers to a very high apparent GRB radiation
efficiency - even higher than the one estimated using the late X-ray
afterglow data (Lloyd-Ronning \& Zhang 2004)\footnote{This could be
attributed to the shallow decay injection phase (segment II in
Fig.\ref{XRTlc} as discussed in \S3.2. Because of the injection, the
effective $E_{iso}$ in the early epochs is smaller than that in the
later epochs. As a result, a larger
$F_{\nu,X}^{prompt}-F_{\nu,X}^{ag}$ contrast is expected.}. 
The commonly-invoked internal shock model predicts a low emission
efficiency (e.g. Panaitescu et al. 1999; Kumar 1999). Understanding
such a high apparent radiation efficiency is therefore desirable (see
e.g. Beloborodov 2000; Kobayashi \& Sari 2001). 

3. The common steep-to-shallow transition feature indicates that the
fireball has already been decelerated at the time when the GRB tail
emission fades. Otherwise, one would see an initially rising
lightcurve peaking at the fireball deceleration time. This fact alone
sets a lower limit to the intial 
Lorentz factor of the fireball, since the deceleration time $t_{dec}$
must be earlier than the transition time $t_{b1}$. The numerical
expression is
\be
\Gamma_0 \geq 125 \left(\frac{E_{\gamma,iso,52}}{\eta_\gamma
n}\right)^{1/8} t_{b1,2}^{-3/8} \left(\frac{1+z}{2}\right)^{3/8}, 
\label{Gam0}
\ee
where $E_{\gamma,iso}$ is the isotropic gamma-ray energy (which is an
observable if the redshift $z$ is known), 
$\eta_\gamma = E_{\gamma,iso} / E_{iso}$ is a conversion factor
between the isotropic afterglow energy $E_{iso}$ and $E_{\gamma,iso}$. 
{Throughout the paper, the convention $Q_x=Q/10^x$ is
adopted in cgs units.}
Applying the method to the bursts with
measured $z$ (Chincarini et al. 2005), we get the lower limits og
$\Gamma_0$ for several
bursts (Table \ref{Tab:Gamma}). Given the weak dependence on the
unknown parameters (i.e. $(\eta_\gamma n)^{-1/8}$), we conclude that
the data suggest that GRBs are highly relativistic events with typical
Lorentz factors higher than 100. This is an independent method, as
compared with previous ones using the high energy spectrum
(e.g. Baring \& Harding 1997; Lithwick \& Sari 2001), the reverse shock
data (Sari \& Piran 1999; Wang et al. 2000; Zhang et al. 2003), and
the superluminal expansion of the radio afterglow source image (Waxman
et al. 1998).
\begin{table}
\caption{
Constraints on the initial Lorentz factors of several GRBs.}

\begin{tabular}{ccccc}
\hline\hline  

GRB &  $z$  &  $t_{b1}({\rm s})$  &  $E_{\gamma,iso,52}$\tablenotemark{a} &
$\Gamma_0 $ \\
\hline
050126...   & 1.290\tablenotemark{b}  &  $\sim 110$ &  0.77   & $> 120 (\eta_\gamma n)^{-1/8}$ \\ 
050315...   & 1.949\tablenotemark{c}  &  $\sim 400$ &  2.77   & $> 100 (\eta_\gamma n)^{-1/8}$ \\
050319...   & 3.240\tablenotemark{d}  &  $\sim 400$ &  5.12   & $> 120 (\eta_\gamma n)^{-1/8}$ \\
050401...   & 2.900\tablenotemark{e}  &  $\sim 130$ & 27.49   & $> 220 (\eta_\gamma n)^{-1/8}$ \\
\hline
\end{tabular}
\tablenotetext{a}{Chincarini et al. 2005}
\tablenotetext{b}{Berger et al. 2005}
\tablenotetext{c}{Kelson \& Berger 2005}
\tablenotetext{d}{Fynbo et al. 2005a}
\tablenotetext{e}{Fynbo et al. 2005b}
\label{Tab:Gamma}
\end{table}

\section{Forward shock emission}
\label{sec:FS} 

After the rapid fading of the GRB tail emission, usually the forward
shock emission component gives the main contribution to 
the early X-ray afterglow lightcurves. The lightcurve shape depends on
the density profile of the ambient medium (i.e. ISM or wind).
In the ``standard'' case (i.e. adiabatic evolution with promp
injection of energy), the fireball energy is
essentially constant during the deceleration phase. The bulk
Lorentz factor $\Gamma \propto R^{-3/2}$ for the ISM case and 
$\Gamma \propto R^{-1/2}$ for the wind case. When the
bulk Lorentz factor $\Gamma$ is larger than $\theta_j^{-1}$, where
$\theta_j$ is the jet opening angle (or the viewing angle of a
structured jet), the system is simply determined by the ratio of the
isotropic afterglow energy $E_{iso}$ and the ambient density $n$ (or
the $A$ parameter in the wind model). Such a ``normal'' decay phase
corresponds to the segment III in the synthetic lightcurve
(Fig. \ref{XRTlc}). When $\Gamma$ becomes smaller
than $\theta_j^{-1}$, the lightcurve steepens because of the
combination of the jet edge effect and the possible sideways expansion
effect (Rhoads 1999; Sari et al. 1999; Panaitescu \& M\'esz\'aros
1999). The bulk Lorentz factor decreases exponentially with
radius. This is known as a ``jet break'', and the post-break segment
corresponds to the segment IV in Fig. \ref{XRTlc}.

During the early evolution of the fireball, the forward shock may be
continuously refreshed with additional energy. This could be either
because of a continuous operation of the central engine (Dai \& Lu
1998a; Zhang \& M\'esz\'aros 2001; Dai 2004), or because of a power
law distribution of the Lorentz factors in the ejecta that results in
slower ejecta catching up with the decelerated fireball at later times
(Rees \& M\'esz\'aros 1998; Panaitescu et al. 1998; Kumar \& Piran
2000b; Sari \& M\'esz\'aros 2000; Zhang \& M\'esz\'aros 2002b), or
because of the transfering of the Poynting-flux energy to the medium
when a Poynting-flux dominated flow is decelerated (e.g. Zhang \&
Kobayashi 2005). The cannonical XRT lightcurve (Fig.\ref{XRTlc})
indeed shows a shallow decay phase (segment II), which we argue is
due to continuous energy injection.

\subsection{Standard afterglow models}
\label{sec:FS1}

For the convenience of the later discussion, we summarize
the ``standard'' early forward shock X-ray afterglow properties as
follows. 

{\em The ISM model} (e.g. Sari et al. 1998). The typical synchrotron
frequency and the cooling frequency are $\nu_m = 6.5\times 10^{14}
~{\rm Hz} ~ \epsilon_{B,-2}^{1/2} \epsilon_{e,-1}^2 E_{52}^{1/2}
t_3^{-3/2} [(1+z)/2]^{1/2}$, and $\nu_c = 2.5 \times 10^{16} ~{\rm Hz}
~ (1+Y)^{-2} \epsilon_{B,-2}^{-3/2} E_{52}^{-1/2} n^{-1} t_3^{-1/2}
[(1+z)/2]^{-1/2}$, respectively, where $E$ is the isotropic kinetic 
energy of the fireball, $n$ is the ISM density, $\epsilon_e$ and
$\epsilon_B$ are shock equipartition parameters for electrons and
magnetic fields, respectively, $Y$ is the energy ratio between the
inverse Compton component and the synchrotron component, $z$ is the
redshift, $t$ is the observer's time. Both frequencies decrease with
time. The time interval for $\nu_m$ and $\nu_c$ to cross the XRT
energy band (0.5-10 keV) from above can be expressed as
\bea
t_m & = & (4-30)~{\rm s}~ \epsilon_{B,-2}^{1/3} \epsilon_{e,-1}^{4/3}
E_{52}^{1/3} \left( \frac{1+z}{2} \right)^{1/3} \label{tm} \\
t_c & = & (0.1-40) ~{\rm s}~ (1+Y)^{-4} \epsilon_{B,-2}^{-3}
E_{52}^{-1} n^{-2} \left( \frac{1+z}{2} \right)^{-1}~. \label{tc}
\eea 
The epoch when the fireball switches from fast cooling ($\nu_c <
\nu_m$) to slow cooling ($\nu_c > \nu_m$) is defined by requiring
$\nu_m=\nu_c$, which reads
\be
t_{mc} = 26 ~{\rm s} ~ (1+Y)^2 \epsilon_{B,-2}^2 \epsilon_{e,-1}^2
E_{52} n \left( \frac{1+z}{2} \right)~.
\label{tmc}
\ee
For comparison, the time when the fireball is decelerated (thin shell
case) is given by 
\bea
t_{dec} & = & \left( \frac{3E} {4\pi n m_p c^2 \Gamma_0^2} \right)^{1/3}
\frac{1}{2 \Gamma_0^2 c} \nonumber \\
& = & 180 ~{\rm s}~ (E_{52}/n)^{1/3}
\Gamma_{0,2}^{-8/3} \left( \frac{1+z}{2} \right)~,
\label{tdec}
\eea
where $\Gamma_0$ is the initial Lorentz factor of the fireball.
We can see that for typical parameters, the XRT band is already in the
regime of $\nu_X > {\rm max} (\nu_m, \nu_c)$ when deceleration
starts. Also, the blast wave evolution has usually entered the slow
cooling regime where the radiative losses are not
important\footnote{In certain parameter regimes, the condition
$t_{dec} < t_{mc}$ could be 
satisfied, and in the temporal regime $t<t_{mc}$, the blastwave is in
the fast cooling ($\nu_c < \nu_m$) domain, so the radiative loss could 
affect the blastwave dynamics (e.g. B\"ottcher \& Dermer
2000; Wu et al. 2005).}.  Under such conditions, the temporal index
is\footnote{Here and below, the convention $F_\nu(X) \propto t^{-\alpha_X} 
\nu^{-\beta_X}$ is adopted since in the X-ray band both the temporal
and the spectral indices are negative. Also $p=2.2$ is adopted for
typical numerical values.} $\alpha_X=(3p-2)/4 \sim
1.15$, and the
spectral index is $\beta_X= p/2 \sim 1.1$ (photon index 2.1). The
relation between the two indices is $\alpha_X=(3 \beta_X-1)/2$. 
In view that $t_c$ is very sensitive to $\epsilon_B$, one can 
still get the regime $\nu_m < \nu_X < \nu_c$ if $\epsilon_B$ is large
(say, around 0.1). In such a case, $\alpha_X=3(p-1)/4 \sim 0.9$,
$\beta_X = (p-1)/2 \sim 0.6$, and $\alpha_X = (3/2) \beta_X$. The
above two cases have been commonly observed in early X-ray afterglows
of many Swift bursts (e.g. segment III in Fig. \ref{XRTlc}),
suggesting that the fireball shock model can successfully interpret
the general properties of GRB afterglows, and that most GRBs are born
in a constant density medium. This conclusion is consistent with
previous analyses (Panaitescu \& Kumar 2002; Yost et al. 2003). 

{\em The wind model} (e.g. Chevalier \& Li 2000). The typical
synchrotron frequency and the cooling frequency are $\nu_m = 1.3\times
10^{15} ~{\rm Hz} ~ \epsilon_{B,-2}^{1/2} \epsilon_{e,-1}^2 E_{52}^{1/2}
t_3^{-3/2} [(1+z)/2]^{1/2}$ (notice that $\nu_m$ in the wind case has
the same parameter dependences as the ISM case but is larger
by a factor of $\sim 2$, e.g. Dai \& Lu 1998b), and $\nu_c = 6.8 \times
10^{12} ~{\rm Hz} ~ (1+Y)^{-2} \epsilon_{B,-2}^{-3/2} E_{52}^{1/2}
A_*^{-2} t_3^{1/2} [(1+z)/2]^{-3/2}$, where $A_*= (\dot M_W/4\pi V_W)
/ (5\times 10^{11} ~{\rm g~cm^{-1}})$ is the typical wind parameter, 
$\dot M_W$ is the mass loss rate, and $V_W$ is the wind velocity. 
A distinguishing property of the wind model is that $\nu_c$ increases
with time. Similarly, the time interval for $\nu_m$ to cross the XRT
band from above is
\be
t_m  =  (6.6-49)~{\rm s}~ \epsilon_{B,-2}^{1/3} \epsilon_{e,-1}^{4/3}
E_{52}^{1/3} \left( \frac{1+z}{2} \right)^{1/3} \label{tm2}~, 
\ee
and the time interval for $\nu_c$ to cross the band from below is
\bea
t_c & = & (3.1\times 10^{11} - 1.2\times 10^{14}) ~{\rm s}~ (1+Y)^4
\epsilon_{B,-2}^3 E_{52}^{-1} A_*^4 \left( \frac{1+z}{2} \right)^{3}
\nonumber \\ 
& = & (0.3 - 120) ~{\rm s}~ (1+Y)^4 \epsilon_{B,-2}^3 E_{52}^{-1}
A_{*,-3}^4 \left( \frac{1+z}{2} \right)^{3} ~. 
\eea
The critical time for the fast-slow cooling transition is
\bea
t_{mc} & = & 1.4\times 10^4 ~{\rm s}~ (1+Y) \epsilon_{e,-1}
\epsilon_{B,-2} A_* \nonumber \\
& = & 14 ~{\rm s}~ (1+Y) \epsilon_{e,-1}
\epsilon_{B,-2} A_{*,-3} ~.
\eea
The deceleration time is essentially the duration of the burst, i.e. 
$t_{dec} \sim T = 20 ~{\rm s}~ T_{1.3}$, since for typical parameters,
the wind model is the so-called ``thick shell'' case (e.g. Chevalier
\& Li 2000; Kobayashi \& Zhang 2003b). For a typical
wind parameter $A_* \sim 1$, the X-ray lightcurve is very simple. For
$t<t_m$, one has $\nu_c < \nu_X < \nu_m$, so that $\alpha_X = 1/4$,
$\beta_X = 1/2$ and $\alpha_X = (1 - \beta_X)/2$. When $t > t_m$,
during most of the observational time of interest, one has $\nu_X >
{\rm max} (\nu_m, \nu_c)$, so that $\alpha_X = (3p-2)/2 \sim 1.15$, 
$\beta_X = p/2 = 1.1$ (photon index 2.1), and $\alpha_X=(3 \beta_X
-1)/2$.  The switching between the fast cooling and slow cooling
regimes does not influence the temporal and spectral indices in the
X-ray band. Only when $A_* < 0.01$, i.e. $t_c$ falls into the range of
observational interest, does a new temporal/spectral domain appear.
When $t>t_c$, one has $\nu_m < \nu_X < \nu_c$, $\alpha_X = (3p-1)/4
\sim 1.4$, $\beta_X = (p-1)/2 \sim 0.6$ (photon index 1.6), and 
$\alpha_X=(3 \beta_X + 1)/2$. Such a feature has been used to
interpret GRB 050128 (Campana et al. 2005)\footnote{As discussed in
\S\ref{sec:FS2}, after collecting more data, we now believe that the
shallow-to-normal decay observed in GRB 050128 is more likely due to
the transition from the energy injection phase to the standard phase
(without injection).}. If $A_*$ is not much smaller than unity,
the blastwave is in the fast cooling regime, and radiative losses
could be substantial (B\"ottcher \& Dermer 2000). A detailed analysis
has been presented in Wu et al. (2005).

{\em The jet model} (e.g. Rhoads 1999; Sari et al. 1999).
After the jet break, the temporal decay index is predicted to be
$\alpha_X = p$. This is derived by assuming significant
sideways expansion. This result is independent on whether the 
X-ray band is below or above $\nu_c$, and whether the medium is an ISM
or a stellar wind. For the latter, the time scale for the lightcurve
to achieve the asymptotic $-p$ index is typically longer than that in
the ISM case (e.g. Kumar \& Panaitescu 2000b; Gou et al. 2001). 

All the above discussions apply for the case of $p>2$. For $p<2$, the
case could be different. Dai \& Cheng (2001) proposed one scenario to
deal with the case of $p < 2$, while Panaitescu \& Kumar (2002)
extended the treatment of $p>2$ case to the $p<2$ regime.

\subsection{Refreshed shock models}
\label{sec:FS2}

If there is significant continuous energy injection into the fireball
during the deceleration phase, the forward shock keeps being
``refreshed'', so that it decelerates less rapidly than in
the standard case. The bulk Lorentz factor of the fireball decays more
slowly than $\Gamma \propto R^{-3/2} (R^{-1/2})$ for the ISM case 
(the wind case), respectively.

There are three possible physical origins for the refreshed shocks.

1. The central engine itself is longer lasting, e.g. behaving as 
\be
L(t) = L_0 (t/t_b)^{-q}~.
\label{Lt}
\ee
The dynamical evolution and the radiation signature of such a system
has been discussed in detail in Zhang \& M\'esz\'aros (2001). A
specific model for such an injection case, i.e. the energy injection
from the initial spin down from a millesecond pulsar (preferably a
millisecond magnetar) was discussed in that paper and earlier in
Dai \& Lu (1998a). In such a specific model, $q=0$ is required
according to the spin-down law. Alternatively, the continued engine
activity could be due to continued infall onto a central black hole,
resulting in the time dependence eq.(\ref{Lt})\footnote{The black hole
- torus system typically has $q=5/3$ at later times (MacFadyen et
al. 2001; Janiuk et al. 2004), which has no effect on the blastwave
evolution.}. In general, for an adiabatic 
fireball, the injection would modify the blastwave dynamics as long as
$q<1$ (Zhang \& M\'esz\'aros 2001). The energy in the fireball
increases with time as $E_{iso} \propto\ t^{1-q}$, so that 
\begin{eqnarray}
\Gamma \propto R^{-\frac{2+q}{2(2-q)}} \propto t^{-\frac{2+q}{8}}, & R\propto
t^{\frac{2-q}{4}}, & {\rm
ISM} \\
\Gamma \propto R^{-\frac{q}{2(2-q)}} \propto t^{-\frac{q}{4}}, & R \propto
t^{\frac{2-q}{2}}, & {\rm wind} 
\end{eqnarray}
It is then straightforward to work out the temporal indices for
various temporal regimes.

{\it The ISM model}. The typical synchrotron
frequency $\nu_m \propto 
\Gamma^2 \gamma_e B \propto \Gamma^4 \propto t^{-(2+q)/2}$, the
synchrotron cooling frequency $\nu_c \propto \Gamma^{-1} B^{-3} t^{-2}
\propto \Gamma^{-4} t^{-2} \propto t^{(q-2)/2}$, and the peak flux
density $F_{\nu,max} \propto N_e B \Gamma \propto t^{1-q}$,
where $B \propto \Gamma$ is the comoving 
magnetic field strength, $\gamma_e \propto \Gamma$ is the typical
electron Lorentz factor in the shocked region, and $N_e \propto R^3$
is the total number of the emitting electrons. The temporal indices
$\alpha$ for various spectral regimes and their relationships with the
spectral indices $\alpha(\beta)$ are listed in Table
\ref{Tab:alpha-beta}. 

{\it The wind model}. In the wind case,
the ambient density is $n\propto R^{-2}$, where $R$ is the radial
distance of the shock front to the central source. The typical
synchrotron frequency $\nu_m \propto \Gamma^2 \gamma_e B \propto
\Gamma^3 B \propto t^{-(2+q)/2}$, the 
synchrotron cooling frequency $\nu_c \propto \Gamma^{-1} B^{-3} t^{-2}
\propto \Gamma^{-4} t^{-2} \propto t^{(2-q)/2}$, and the peak flux
density $F_{\nu,max} \propto N_e B \Gamma \propto \Gamma^2 \propto
t^{-q/2}$, where $B \propto \Gamma R^{-1}$ is the 
comoving magnetic field strength, and $N_e \propto R$ is the total
number of the emitting electrons. The 
temporal indices $\alpha$ for various spectral regimes and their
relationships with the spectral indices $\alpha(\beta)$ are listed in
Table \ref{Tab:alpha-beta}. 

In order for the central engine to continuously feed the
blast wave, the Lorentz factor of the continuous flow must be  
(much) larger than that of the blast wave. It could be a 
Poynting-flux-dominated flow. This is not difficult to satisfy 
since the blast wave keeps decelerating. There could be a reverse
shock propagating into the continuous ejecta, but the radiation
signature  of the reverse shock is typically not in the X-ray band
(e.g. Zhang \& M\'esz\'aros 2001).

2. The central engine activity may be brief (e.g. as brief as the
prompt emission itself, but at the end of the prompt phase, the ejecta
has a range of Lorentz factors, e.g., the amount of ejected mass
moving with Lorentz factors greater than $\gamma$ is (Rees \&
M\'esz\'aros 1998; Panaitescu et al. 1998; Sari \& M\'esz\'aros 2000)
\be
M(>\gamma) \propto \gamma^{-s}~.
\ee
The total energy in the fireball increases as $E_{iso} \propto
\gamma^{1-s} \propto \Gamma^{1-s}$, so that
\begin{eqnarray}
\Gamma \propto R^{-\frac{3}{1+s}} \propto t^{-\frac{3}{7+s}}, & R\propto
t^{\frac{1+s}{7+s}}, & {\rm ISM} \\
\Gamma \propto R^{-\frac{1}{1+s}} \propto t^{-\frac{1}{3+s}}, & R \propto
t^{\frac{1+s}{3+s}}, & {\rm wind} 
\end{eqnarray}
One can then work out the temporal decay indices in various spectral
regimes (e.g. Rees \& M\'esz\'aros 1998; Sari \& M\'esz\'aros 2000). 
Alternatively, for each $s$ value, one can find an effective $q$
value that mimics the $s$ effect, or vice versa. This gives
\begin{eqnarray}
s =\frac{10-7q}{2+q}, ~~ q=\frac{10-2s}{7+s}, & {\rm ISM} \label{s-q-1}\\
s =\frac{4-3q}{q}, ~~ q=\frac{4}{3+s}. & {\rm wind} \label{s-q-2}
\end{eqnarray}
In Table \ref{Tab:alpha-beta}, the explicit $s$-dependences are not
listed, but they could be inferred from eqs.(\ref{s-q-1}) and
(\ref{s-q-2}). 

In this second scenario, the central engine need not last
long. All the material could be ejected promptly. The continuous
injection is due to the different velocities of the ejecta. Initially 
as the blast wave moves with high speed, the slower ejecta lag 
behind and have no effect on the blastwave evlolution. They later 
progressively pile up onto the blast wave as the latter decelerates. 
Only when $s > 1$ does one expect a change in the fireball dynamics.
This corresponds to $q<1$. For $q=0.5$, one gets $s=2.6$ for the ISM case 
and $s=5$ for the wind case.
\clearpage
\begin{table}
\caption{Temporal index $\alpha$ and spectral index
$\beta$ in various afterglow models.\label{Table1}} 
\tabletypesize{\scriptsize}
\begin{tabular}{llllll}
\hline\hline  
& & no injection & &  injection & \\

& $\beta$ & $\alpha $  &  $\alpha (\beta)$ & $\alpha$ & $\alpha (\beta)$  \\
\hline
ISM & slow cooling \\
\hline
$\nu<\nu_m$   &  $-{1 \over 3}$  &   $-{1\over 2}$ & $\alpha={3\beta \over 2}$ & ${5q-8 \over 6}$ (-0.9) & $\alpha=(q-1)+\frac{(2+q)\beta}{2}$\\
$\nu_m<\nu<\nu_c$   &  ${{p-1 \over 2}}$ (0.65)  &  ${3(p-1)\over 4}$
(1.0)   &  $\alpha={3\beta \over 2}$ & ${(2p-6)+(p+3)q \over 4}$ (0.3) & $\alpha=(q-1)+\frac{(2+q)\beta}{2}$\\
$\nu>\nu_c$   &  ${{p\over 2}}$ (1.15)  &   ${3p-2 \over 4}$ (1.2) & $\alpha={3\beta-1 \over 2}$ & ${(2p-4)+(p+2)q\over 4}$ (0.7) & $\alpha=\frac{q-2}{2}+\frac{(2+q)\beta}{2}$ \\

\hline
ISM & fast cooling \\
\hline
$\nu<\nu_c$   &  $-{1\over 3}$  &   $-{1\over 6}$ &  $\alpha={\beta \over 2}$  &  ${7q-8 \over 6}$ (-0.8) & $\alpha=(q-1)+\frac{(2-q)\beta}{2}$\\
$\nu_c<\nu<\nu_m$   &  ${1\over 2}$  &  ${1\over 4}$  & $\alpha={\beta \over 2}$ &  ${3q-2 \over 4}$ (-0.1)  &  $\alpha=(q-1)+\frac{(2-q)\beta}{2}$ \\
$\nu>\nu_m$   &  ${p\over 2}$ (1.15)  &   ${3p-2\over 4}$ (1.2)     &  $\alpha={3\beta-1 \over 2}$  &  ${(2p-4)+(p+2)q\over 4}$ (0.7) & $\alpha=\frac{q-2}{2}+\frac{(2+q)\beta}{2}$ \\

\hline
Wind & slow cooling \\
\hline
$\nu<\nu_m$   &  $-{1\over 3}$   &   0  &  $\alpha={3\beta+1 \over 2}$ & ${q-1 \over 3}$ (-0.2) & $\alpha=\frac{q}{2}+\frac{(2+q)\beta}{2}$\\
$\nu_m<\nu<\nu_c$   &  ${p-1\over 2}$ (0.65)  &   ${3p-1\over 4}$ (1.5)   &   $\alpha={3\beta+1 \over 2}$ & ${(2p-2)+(p+1)q \over 4}$ (1.1)& $\alpha=\frac{q}{2}+\frac{(2+q)\beta}{2}$\\
$\nu>\nu_c$   &  ${p\over 2}$ (1.15)  &   ${3p-2\over 4}$ (1.2)   &  $\alpha={3\beta-1 \over 2}$   &   ${(2p-4)+(p+2)q\over 4}$ (0.7) &  $\alpha=\frac{q-2}{2}+\frac{(2+q)\beta}{2}$\\

\hline
Wind & fast cooling  \\
\hline
$\nu<\nu_c$   &  $-{1\over 3}$   &   ${2\over 3}$ & $\alpha={1-\beta \over 2}$ &  ${(1+q) \over 3}$ (0.5) & $\alpha=\frac{q}{2}-\frac{(2-q)\beta}{2}$  \\
$\nu_c<\nu<\nu_m$   &  ${1\over 2}$  &   ${1\over 4}$  & $\alpha={1-\beta \over 2}$  & ${3q-2 \over 4}$ (-0.1)& $\alpha=\frac{q}{2}-\frac{(2-q)\beta}{2}$\\
$\nu>\nu_m$   &  ${p\over 2}$ (1.15) &   ${3p-2\over 4}$ (1.2)    & $\alpha={3\beta-1 \over 2}$  &  ${(2p-4)+(p+2)q\over 4}$ (0.7) & $\alpha=\frac{q-2}{2}+\frac{(2+q)\beta}{2}$\\
\hline
\end{tabular}
\label{Tab:alpha-beta}
\tablecomments{This is the extension of the
Table 1 of Zhang \& M\'esz\'aros (2004), with the inclusion of the
cases of energy injection. The case of $p<2$ is not included, and
the self-absorption effect is not discussed. Notice
that a different convention $F_\nu \propto t^{-\alpha} \nu^{-\beta}$
is adopted here (in comparison to that used in Zhang \& M\'esz\'aros
2004), mainly because both the temporal index and the
spectral index are generally negative in the X-ray band. The temporal
indices with energy injection are valid only for $q < 1$, and they
reduce to the standard case (without energy injection, e.g. Sari et
al. 1998, Chevalier \& Li 
2000) when $q=1$. For $q>1$ the expressions are no longer valid, and
the standard model applies. An injection case due to pulsar spindown
corresponds to $q=0$ (Dai \& Lu 1998a; Zhang \& M\'esz\'aros
2001). Recent {\em Swift} XRT data are generally consistent with $q
\sim 0.5$. The numerical values quoted in parentheses are for $p=2.3$
and $q=0.5$. }
\end{table}
\clearpage

3. The energy injection is also brief, but the outflow has a
significant fraction of Poynting flux (e.g. Usov 1992; Thompson 1994;
M\'esz\'aros \& Rees 1997b; Lyutikov \& Blandford 2005). Assigning 
a parameter $\sigma$ for the outflow, which is the ratio between the
Poynting flux and baryonic kinetic energy flux, Zhang \& Kobayashi
(2005) modeled the reverse shock emission from ejecta with an
arbitrary $\sigma$ value. They found that during the crossing of the
reverse shock, the Poynting energy is not transferred to the ambient
medium.  The Poynting energy (roughly by a factor of $\sigma$) is
expected to be transferred to the medium (and hence, to the afterglow
emission) after the reverse shock disappears. Zhang \& Kobayashi
(2005) suggest that the transfer is delayed with respect to the
traditional case of $\sigma=0$. The energy transfer process, however,
is poorly studied so that one does not have a handy conversion
relation with the $q$ value derived in the first scenario. 

\subsection{Case studies}

In this subsection, we discuss several {\em Swift} GRBs with
well-monitored early afterglow data detected by XRT. The notations for
the break times and the temporal slopes are per those marked on
Fig. \ref{XRTlc}. 

{\bf GRB 050128} (Campana et al. 2005): The lightcurve can be fitted
by a broken power law with the break time at $t_{b2} =
1472^{+300}_{-290}$s. The temporal decay indices before and after the
break are $\alpha_2=0.27^{+0.12}_{-0.10}$ and
$\alpha_3=1.30^{+0.13}_{-0.18}$, respectively. The spectral indices
before and after the break are essentially unchanged, i.e. $\beta_2
\sim 0.59\pm 0.08$, and $\beta_3 = 0.79\pm 0.11$. Campana et
al. (2005) discussed two interpretations. A jet model requires a very
flat electron spectral index, i.e. $p \sim 1.3$, as well as a change
of the spectral domain before and after the jet break. Alternatively,
the data may be accommodated in a wind model, but one has to assume
three switches of the spectral regimes during the observational gap
from 400s to about 4000s. So neither explanation is completely
satisfactory. By comparing the predicted indices in Table
\ref{Tab:alpha-beta}, the observation may be well-interpreted within
the ISM model with an initial continuous energy injection episode. The
segment after 
the break is consistent with a standard ISM model for $\nu_m < \nu_X <
\nu_c$, with $p \sim 2.6$. The lightcurve before the break, on the
other hand, is consistent with an injection model with $p \sim 2.2$
and $q \sim 0.5$ in the same spectral regime. The break time is
naturally related to the cessation of the injection process, and a
slight change of electron spectral index (from 2.2 to 2.6) is
required. From the beginning of the observation (100s) to $t_{b2}$,
the total energy is increased by a factor of $(1472/100)^{(1-0.5)}
\sim 2.8$. 

{\bf GRB 050315} (Vaughan et al. 2005): After a steep decay ($\alpha_1
= 5.2^{+0.5}_{-0.4}$) up to $t_{b1}=308$s, the lightcurve shows a flat
``plateau'' with a temporal index of $\alpha_2
=0.06^{+0.08}_{-0.13}$. It then turns to $\alpha_3=0.71\pm 0.04$ at
$t_{b2}=1.2^{+0.5}_{-0.3}\times 10^4$s. Finally there is a third break
at $t_{b3}=2.5^{+1.1}_{-0.3}\times 10^5$s, after which the temporal
decay index is $\alpha_4=2.0^{+1.7}_{-0.3}$. So this burst displays
all four segments presented in Fig.\ref{XRTlc}. The spectral indices
in segments II, III and IV are essentially constant,
i.e. $\beta_2=0.73\pm 0.11$, $\beta_3=0.79\pm 0.13$ and
$\beta_4=0.7^{+0.5}_{-0.3}$, respectively. Segment III is
consistent with an ISM model with  
$\nu_X > \nu_c$ and $p=1.6$, since in this model $\beta=p/2=0.8$,
$\alpha=(3p-2)/4=0.7$, in perfect agreement with the data. The
third temporal break $t_{b3}$ is consistent with a jet
break. According to Dai \& Cheng (2001), the post-break temporal index
for $p<2$ is $\alpha=(p+6)/4=1.9$, which is also consistent with the
observed $\alpha_4$. The plateau between $t_{b1}$ and $t_{b2}$ is then
due to an energy injection in the same ISM model ($\nu_X > \nu_c$),
with $p \sim 1.5$ and $q \sim 0.35$. The total injected energy is
increased by a factor of $(12000/308)^{(1-0.35)} \sim 11$.

{\bf GRB 050319} (Cusumano et al. 2005): After a
steep decay ($\alpha_1 = 5.53\pm 0.67$) up to $t_{b1}=(384\pm 22)$s, the
lightcurve shows a shallow decay with a temporal index of $\alpha_2
=0.54\pm 0.04$. It steepens to $\alpha_3=1.14\pm 0.2$ at
$t_{b2}=(2.60\pm 0.70) \times 10^4$s. The spectral indices in
segment II and III are $\beta_2=0.69\pm 0.06$ and $\beta_3=0.8\pm 0.08$,
respectively. Again segment III is well consistent with an ISM model
for $\nu_m < \nu_X < \nu_c$ with $p=2.6$, which gives
$\beta=(p-1)/2=0.8$ and $\alpha=(3/2)\beta=1.2$, in excellent
agreement with the data. Interpreting the segment II (the shallow decay
phase) as the energy injection phase, for the same ISM model ($\nu_m <
\nu_X < \nu_c$), one gets $p\sim 2.4$ and $q \sim 0.6$. The total
injected energy is increased by a factor of $(26000/384)^{(1-0.6)} \sim
5.4$. The UVOT observations are also consistent with such a picture
(Mason et al. 2005). 

{\bf GRB 050401} (de Pasquale et al. 2005): The early X-ray lightcurve
is consistent with a broken power law, with $\alpha_2=0.63 \pm 0.02$,
$\alpha_3 = 1.41 \pm 0.1$, and $t_{b2}=4480^{+520}_{-440}$s. The
spectral indices before and after the break are all consistent with
$\beta_2 \sim \beta_3 = 0.90 \pm 0.03$. The $\alpha-\beta$ relation
does not fit into a simple $p<2$ jet model. 
On the other hand, the energy injection model gives a natural
interpretation. After the 
break, the lightcurve is consistent with an ISM model for $\nu_m <
\nu_X <\nu_c$ with $p=2.8$. Before the break, it is consistent with
the same model with $q=0.5$. The total injected energy is increased by
a factor of $>(4480/200)^{(1-0.5)} \sim 4.7$.

The injection signature is also inferred in other bursts such as GRB
050117 (Hill et al. 2005) and XRF 050416 (Sakamoto et al. 2005),
where similar conclusions could be drawn. The injection model is
supported by an independent study of Panaitescu et al. (2005).

\subsection{Theoretical implications}

The following conclusions could be drawn from the above case studies.

1. A common feature of the early X-ray afterglow lightcurves is a well
defined temporal steepening break. A crucial observational
fact is that there is essentially no spectral variation before and 
after the break. This suggests that the break is of hydrodynamic 
origin rather than due to the crossing of some typical frequencies of
the synchrotron spectrum in the band. It is worth mentioning that a
lightcurve transition similar to the transition between segments II
and III is expected in a radiative fireball (e.g. B\"ottcher \& Dermer
2000), see e.g. Figs. 1 \& 2 of Wu et al. (2005). However, that
transition is due to the crossing of $\nu_m$ in the observational
band. One therefore expects a large spectral variation before and
after the break, which is inconsistent with the data.
Another straightforward interpretation would
be a jet break, but there are three reasons against such an
interpretation. First, in all the cases, $p<2$ has to be assumed. This
is in stark contrast to the late jet breaks observed in the optical
band, which typically have $p>2$. Furthermore, the $\alpha-\beta$
relation predicted in the jet model is usually not satisfied. Second,
the post-break $\alpha-\beta$ relation is usually satisified in a
standard slow cooling ISM model, with the X-ray band either below or
above the cooling frequency. In such a sense, this segment is quite
``normal''. Third, in some cases (e.g. GRB 050315), another steepening
break is observed after this normal segment, which is consistent with
the jet break interpretation. Since only one break could be attributed
to a jet break, the ``shallow-to-normal'' break must be due to
something else.

2. A natural interpretation of the shallow decay phase is to attribute
it to a continuous energy injection, so that the forward shock is
``refreshed''. Three possibilities exist to account for the refreshed 
shock effect (\S\ref{sec:FS2}): a long-lived central engine with
progressively reduced activities, an instantaneous injection with a
steep power-law distribution of the shell Lorentz factors, and the
deceleration of an instantaneously-injected highly magnetized 
(high-$\sigma$) flow. In terms of afterglow properties, these
possibilities are degenerate (e.g. the connection between
$q$ and $s$) and can not be differentiated. In principle, the first
scenario may give rise to additional observational signatures
(e.g. Rees \& M\'esz\'aros 2000; Gao \& Wei 2004, 2005), which may be
used to differentiate the model from the others.  

3. Two interesting characteristics during the injection phase are that
the injection process is rather smooth, and that the effective $q$
value is around 0.5. This gives interesting constraints on the
possible physical mechanisms. (1) For the scenario of a continuously
injecting central engine (Zhang \& M\'esz\'aros 2001), the central
engine luminosity must vary with time smoothly. This is in contrast
to the conventional GRB central engine which injects energy
erratically to allow the observed rapid variability in the
lightcurves. This usually requires two different energy components,
i.e. one ``hot'' fireball component that leads to the prompt emission
and a ``cold'' Poynting flux component that gives to the smooth
injection. A natural Poynting flux component is due to the spin-down
of a new-born millisecond pulsar (Dai \& Lu 1998a; Zhang \&
M\'esz\'aros 2001). However, a straightforward prediction from such a
model is $q=0$, not consistent with $q \sim 0.5$ inferred from the
observations. Modifications to the simplest model are needed.
Alternatively, the system may be a long-lived black
hole torus system with a reducing accretion rate. However, at later
times the long-term central engine power corresponds to $q=5/3$
(MacFadyen et al. 2001; Janiuk et al. 2004), too steep to give an
interesting injection signature. It is worth mentioning that in the
collapsar simulations (MacFadyen et al. 2001), an extended flat
injection episode sometimes lasts for $\sim 1000$s, which could
potentially interpret the short injection phase of some bursts, but is
difficult to account for some other bursts whose injection phase is
much longer. (2) For the scenario
of a power-law distribution of Lorentz factors (Rees \& M\'esz\'aros
1998), one should require that a smooth distribution of Lorentz
factors is produced after the internal shock phase. In the internal
shock model, slow shells are indeed expected to follow the fast
shells, but they tend to be discrete and give rise to bumpy
lightcurves (e.g. Kumar \& Piran 2000a) especially when the
contribution from the reverse shock is taken into account (Zhang \&
M\'esz\'aros 2002b). It is also unclear how an effective $q \sim 0.5$
is expected. (3) Deceleration of a promptly-ejected Poynting-flux
dominated flow (e.g. Zhang \& Kobayashi 2005) naturally gives a smooth
injection signature observed. Above case studies indicate that the
injected energy is by a factor of several to 10. Within such a
picture, the unknown $\sigma$ value is about several to 10. However,
it is unclear how long the delay would be and whether one can account
for the shallow decay with $q \sim 0.5$ extending for $10^4$ seconds. 
Further more detailed theoretical modeling is needed to test this
hypothesis.  

4. Any model needs to interpret the sudden cessation of the injection
at $t_{bs}$. This time has different meanings within the three
scenarios discussed above. (1) Within the long-lived
central engine model, this is simply the epoch when the injection
process ceases. In the pulsar-injection model, there is a well-defined
time for injection to become insignificant (Dai \& Lu 1998a; Zhang
\& M\'esz\'aros 2001), but within a black-hole-torus injection model,
such a time is not straightforwardly defined. (2) In the varying
Lorentz factor scenario, 
this time corresponds to a cut-off of the Lorentz factor distribution
at the low end below which the distribution index $s$ is flatter than
1 so that they are energetically unimportant. This lowest Lorentz
factor is defined by
\be
\Gamma_m = 23 \left(\frac{E_{iso,52}}{n}\right)^{1/8}
t_{b2,4}^{-3/8} \left(\frac{1+z}{2}\right)^{3/8}, 
\ee
A successful model must be able to address a well-defined
$\Gamma_m$ in this model. (3) Within the Poynting flux
injection model, a well-defined time cut-off
is expected, which corresponds to the epoch when all the Poynting
energy is transferred to the blastwave. If the shallow decay is indeed
due to Poynting energy transfer, the cut-off time ($t_{b2}$) could be
roughly defined by the $\sigma$ parameter through
$\sigma \sim ({t_{b2}}/{t_{dec}})^{(1-q)}$,
where $t_{dec}$ is the conventional deceleration time
defined by $E_{iso}/(1+\sigma)$, when only a fraction of
$(1+\sigma)^{-1}$ energy is transfered to the ISM (Zhang \& Kobayashi
2005).

5. Although we have not tried hard to rule out a wind-model
interpretation, the case studies discussed above suggest that the
early afterglow data are consistent with an ISM model for essentially
all the bursts. This conclusion also applies to other well-studied {\em
Swift} bursts (e.g. GRB 050525a, Blustin et al. 2005). This result is
intriguing given that long GRBs are associated with the death of massive stars,
from which a strong wind is expected. Previous analyses using late
time afterglow data (e.g. Panaitescu \& Kumar 2002; Yost et al. 2003)
have also suggested that most afterglow data are consistent with an ISM
model rather than a wind model. In order to accommodate the data,
it has been suggested that the wind parameter may be small so that at
a late enough time the blastwave is already propagating in an ISM 
(e.g. Chevalier et al. 2004). The {\em Swift} results push the ISM
model to even earlier epochs (essentially right after the
deceleration), and indicate the need for a re-investigation of the
problem. The epoch shortly before the deaths of massive stars
is not well studied (Woosley et al. 2003). One possibility is that the
stellar wind ceases some time before the star collapses. Careful 
analyses of early afterglows of a large sample of long GRBs may shed 
light on the final stage of massive star evolution.

\section{Reverse shock emission}
\label{sec:RS} 

\subsection{Synchrotron emission}
It is generally believed that a short-lived reverse shock exists
during the intial deceleration of the fireball and gives interesting
emission signatures in the early afterglow phase. Given the same
internal energy in both the forward-shocked and the reverse-shocked
regions, the typical synchrotron frequency for the reverse shock
emission is typically much lower than that in the forward shock
region, since the ejecta is much denser than the medium. While the
early forward shock synchrotron emission peaks in X-rays at early
times, the reverse shock synchrotron emission usually peaks in the
optical/IR band or even lower (e.g. M\'esz\'aros \& Rees 1997a; Sari \&
Piran 1999; Kobayashi 2000; Zhang et al. 2003; Zhang \& Kobayashi
2005). This model has been succussful in interpreting the early
optical emission from GRB 990123 (Akerlof et al. 1999; Sari \&
Piran 1999; M\'esz\'aros \& Rees 1999), GRB 021211 (Fox et al. 2003; 
Li et al. 2003; Wei 2003), and GRB 041219a (Blake et al. 2005;
Vestrand et al. 2005; Fan et al. 2005b). As a result, it is expected
that the reverse shock component has a negligible contribution in the
X-ray band.

In the above argument, it has been assumed that the shock parameters
($\epsilon_e$, $\epsilon_B$ and $p$) are the same in both shocks. In
reality this might not be the case. In particular, the GRB outflow may
itself carry a dynamically important magnetic field component (or
Poynting flux). This magnetic field would be shock-compressed and
amplified, giving a larger effective $\epsilon_B$ (Fan et al. 2004a,b;
Zhang \& Kobayashi 2005). Since the medium is generally not
magnetized, it is natural to expect different $\epsilon_B$ values in
both regions, and a parameter ${\cal R}_B \equiv (\epsilon_{B,r} /
\epsilon_{B,f})^2$ have been used in the reverse shock analysis.
It has been found that ${\cal R}_B$ is indeed larger than unity for
GRB 990123 and GRB 021211 (Zhang et al. 2003; Fan et al. 2002; Kumar
\& Panaitescu 2003; Panaitescu \& Kumar 2004; MaMahon et
al. 2004). Hereafter the subscript/superscript ``f'' and ``r''
represent the forward shock and the reverse shock,
respectively. According to Zhang \& Kobayashi (2005), the case of GRB
990123 corresponds to the most optimized case with $\sigma \sim 1$, so
that ${\cal R}_B$ is the largest. 

More generally, $\epsilon_e$ and $p$ may also vary in both shocks. 
Fan et al. (2002) performed a detailed fit to the GRB 990123 data and
obtained  $\epsilon_e^{\rm r}=4.7\epsilon_e^{\rm f}$ and 
$\epsilon_B^{\rm r}=400\epsilon_B^{\rm f}$. A general treatment
therefore requires that we introduce one more parameter, i.e. ${\cal
R}_{e}=[(p^{\rm r}-2)/(p^{\rm r}-1)]/[(p^{\rm f}-2)/(p^{\rm
f}-1)] (\epsilon_{\rm e}^{\rm r}/\epsilon_{\rm e}^{\rm f})$. 
Following the treatment of Zhang et al. (2003), we have the following
relations in the thin-shell regime (see also Fan \& Wei
2005)\footnote{For an
arbitrary $\sigma$, the treatment becomes more complicated. The
treatment presented here is generally valid for $\sigma \lesssim 1$. For
$\sigma > 1$, the reverse shock emission starts to be suppressed
(Zhang \& Kobayashi 2005). Since we are investigating the most
optimistic condition for the reverse shock contribution, in this paper
we adopt the standard hydrodynamic treatment which is valid for $\sigma
\lesssim 1$.} 
\bea
\frac{\nu_{\rm m}^{\rm r}(t_{\rm \times})}
{\nu_{\rm m}^{\rm f}(t_{\rm \times})}& \sim &{\cal R}_{\rm B} 
{\cal R}_{\rm e}^2 \left(\frac{\gamma_{34,\times}-1} 
{\Gamma_{\times}-1}\right)^2,\label{Rnum}\\
\frac{\nu_{\rm c}^{\rm r}(t_\times)}{\nu_{\rm c}^{\rm f}(t_\times)}
& \sim & {\cal R}_{\rm B}^{-3} \left(\frac{1+Y^{\rm f}}
{1+Y^{\rm r}}\right)^2, \label{Rnuc}\\
\frac{F_{\rm \nu, max}^{\rm r}(t_{\rm \times})}
{F_{\rm \nu, max}^{\rm f}(t_{\rm \times})}
& \sim & {\cal R}_{\rm B} \frac{\Gamma_\times^2}{\Gamma_0} \sim {\cal
R}_B \Gamma_0~ \label{RFnum},
\eea
where $Y^{\rm f}$ and $Y^{\rm r}$ are the Compton parameters for the
forward and the reverse shock emission components, respectively; 
$t_\times$ is the reverse shock crossing time, which is essentially
$t_{dec}$ (eq.[\ref{tdec}]) for the thin shell case; $\Gamma_\times$
is the bulk Lorentz factor of the outflow at $t_\times$; 
$\gamma_{34,\times}\approx
(\Gamma_0/\Gamma_\times+\Gamma_\times/\Gamma_0)/2$ is the Lorentz
factor of the shocked ejecta relative to the unshocked one. 

For typical parameters, both $\nu_m^{\rm f}$ and 
 $\nu_c^{\rm f}$ are below the XRT band (comparing eqs.[\ref{tm}],
[\ref{tc}] with eq.[\ref{tdec}]). According to eqs.(\ref{Rnum}) and
(\ref{Rnuc}), $\nu_c^{\rm r}$, $\nu_m^{\rm r}$ should be also below
the XRT band. Following the standard synchrotron emission model, we
then derive the X-ray flux ratio of the reverse shock and the forward
shock components at $t_\times$:
\be
{F_{\nu,X}^{\rm r}(t_\times) \over F_{\nu,X}^{\rm f}(t_\times)}
\approx  {\cal R}_B^{\frac{p-2}{2}} {\cal R}_e^{p-1} \Gamma_0
\left(\frac{\gamma_{34, \times}-1}{\Gamma_\times -1} \right)^{p-1}
\left({1+Y^{\rm f}\over 1+Y^{\rm r}}\right).
\ee 
We can see that for ${\cal R}_B={\cal R}_e=1$, $Y^{\rm f}=Y^{\rm
r}$, and $p \geq 2$, one has $F_{\nu,X}^{\rm r}(t_\times) \lesssim
F_{\nu,X}^{\rm f}(t_\times)$, since $\gamma_{34,\times} \lesssim 1$,
and $\Gamma_\times \sim \Gamma_0$ in the thin shell case. The reverse
shock contamination in the X-ray band is therefore not important. The
situation changes if we allow higher ${\cal R}_e$ and ${\cal R}_B$
values. Increasing ${\cal R}_e$ directly increases the
reverse-to-forward flux ratio. Although the dependence on ${\cal R}_B$
is only mild when $p$ is close to 2, a higher ${\cal R}_B$
suppresses the IC process in the reverse shock region relative to that
in the forward shock region, so that the ratio $(1+Y^{\rm
f})/(1+Y^{\rm r})$ also increases. As a result, as ${\cal R}_B \gg 1$
and ${\cal R}_e \gg 1$, the reverse shock synchotron component would
stick out above the forward shock synchrotron component, and an X-ray
bump is likely to emerge (see also Fan \& Wei 2005). As a numerical example,
taking $p=2.3$, $\Gamma_\times \approx \Gamma_0/2=50$, ${\cal R}_B=10$,
${\cal R}_e=5$, and $\epsilon_e^{\rm f}=30\epsilon_B^{\rm f}$, we get
${F_{\nu_X}^{\rm r}(t_\times) / F_{\nu_X}^{\rm f}(t_\times)}\approx
6$. This could potentially explain the X-ray flare (by a factor of
$\sim 6$) detected in GRB 050406 (Burrows et al. 2005; P. Romano et
al. 2005, in preparation). However, a big
caveat of such a model is that one expects a very bright UV/Optical
flash due to the large ${\cal R}_B$ and ${\cal R}_e$ involved - like
the case of GRB 990123. Unless this flash is completely suppressed by
extinction, the 
non-detection of such a flash in the UVOT band for GRB 050406 strongly
disfavors such an interpretation. 

\subsection{Synchrotron self-Compton emission}

The synchrotron photons in the reverse shock region will be scattered
by the same electrons that produce these photons. The characteristic
energy of this component is typically in the $\gamma$-ray range.
However, under some conditions, this
synchrotron self-Compton (SSC) component would also stand out in the
X-ray band, giving rise to an X-ray bump in the lightcurve. A detailed
discussion has been presented in Kobayashi et al. (2005), which we do
not repeat here. The general conclusion is that the SSC component could
account for an early X-ray flare bump by a factor of several under
certain optimized conditions. An advantage of this model over the
reverse shock synchrotron model is that a bright UV-optical flash is
avoided. However, this model can not account for a flare with a very
large contrast (e.g. by a factor of 500, as seen in GRB 050502B, Burrows
et al. 2005). 

\section{Mechanisms to produce early X-ray flares}
\label{sec:flare} 

XRT observations indicate that X-ray flares are common features in the
early phase of X-ray afterglows (the component V in Figure
\ref{XRTlc}). After the report of the first two cases of flares in GRB
050406 and GRB 050502B (Burrows et al. 2005), 
later observations indicate that nearly half of long GRBs harbor early
X-ray flares (e.g. O'Brien et al. 2005). More intriguingly, the early
X-ray afterglow of the 
latest localized short GRB 050724 (Barthelmy et al. 2005b) also
revealed flares similar to those in the long GRBs (e.g. 050502B). The common
feature of these flares is that the decay indices after the flares are
typically very steep, with a $\delta t/t$ much smaller than
unity. In some cases (e.g. GRB050724, Barthelmy et al. 2005), the
post-flare decay slopes are as steep as $\leq -7$. In this
section we discuss various possible models to interpret the flares and
conclude that the data require that the central engine is active and
ejecting these episodic flares at later times.


\subsection{Emission from the reverse shock?}
\label{subsec:model2}

As discussed in \S\ref{sec:RS}, synchrotron or SSC emission from the
reverse shock region could dominate that from the forward shock
emission in the X-ray band under 
certain conditions. Because of the lack of strong UV-optical flares in
the UVOT observations, we tentatively rule out the reverse shock
synchrotron emission model. The prediction of the lightcurve in the
SSC model could potentially interpret the X-ray flare seen in GRB
050406 (Burrows et 
al. 2005), but the predicted amplitude is too low to interpret the
case of GRB 050502B (Burrows et al. 2005). In some bursts (e.g. GRB
050607), more than one flare are seen in a burst. Although one of
these flares may be still interpreted as the reverse shock SSC
emission, an elegant model should interpret these flares by a model
with the same underlying physics. We tentatively conclude that the
reverse shock model cannot account for most of the X-ray flares
detected by the {\em Swift} XRT in a unified manner.

\subsection{Density clouds surrounding the progenitor?}
\label{subsec:model3}

Long-duration GRBs are believed to be associated with the deaths of
massive stars, such as Wolf-Rayet stars. According to Woosley et
al. (2003), there are no observations to constrain the mass loss rate
of a Wolf-Rayet star during the post-helium phase (100-1000 years
before the explosion), No stability analyses have been carried out to
assess whether such stars are stable. At the high end of a reasonable
range of the mass loss rate, dense clouds surrounding a GRB progenitor
may exist. For a wind velocity $v_{\rm w}\sim 100{\rm
km~s^{-1}}$, the density clumps could occur at a radius $\sim 3\times
10^{16}-3\times10^{17}$cm. These density clumps have been invoked by
Dermer \& Mitman (1999, 2003) to interpret the GRB prompt emission
variabilities. 

\begin{figure}
\plotone{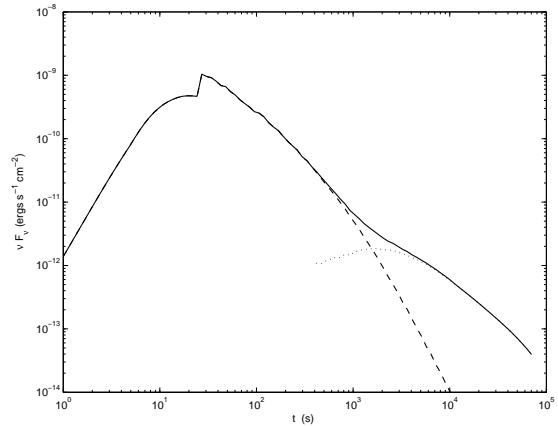}
\caption{The X-ray flare powered by a relativistic fireball
interacting with a dense cloud. The cloud is located at  
$R=(3.0, 3.3)\times 10^{16}$ cm with a density $n_{cloud}=100~{\rm
cm^{-3}}$. The background ISM density is $n=1~{\rm cm^{-3}}$.
Other parameters include: $E_{\rm iso}=10^{52}$ergs, $z=1$,
$\Gamma_0=240$,  $\epsilon_{\rm e}=0.1$, $\epsilon_{\rm B}=0.01$,
$p=2.3$ and $\theta_{\rm j}=0.1$. The dashed line is the X-ray
emission contributed by the electrons shocked at $R<3.3\times
10^{16}$cm, and the dotted line is the X-ray emission contributed by
the electrons shocked at $R>3.3\times 10^{16}$cm. The solid line is
the sum of these two components.}
\label{Fig:Cloud}
\end{figure}

One immediate question is whether these density clumps,
if they exist, could give rise to the X-ray flares detected by XRT.
In order to check this possibility, we investigate the following toy
model. For simplicity, we assume a dense clump extends from $R\sim
3\times 10^{16}{\rm cm}$ to $(R+\Delta) \sim 3.3\times 10^{16}{\rm cm}$
with a number 
density $n_{cloud}\sim 100{\rm cm^{-3}}$. The background ISM density
is taken as $n\sim 1~{\rm cm^{-3}}$. Other parameters in the
calculation include: $E_{\rm iso}=10^{52}$ergs, $z=1$,
$\Gamma_0=240$, $\epsilon_{\rm e}=0.1$, $\epsilon_{\rm B}=0.01$,
$p=2.3$ and $\theta_{\rm j}=0.1$. No sideways expansion of the jet is
included, which is consistent with previous simulations (Kumar \&
Granot 2003; Cannizzo et al. 2004). The X-ray lightcurve is shown in 
Fig.\ref{Fig:Cloud}, the general feature of which is reproduced in
both codes used in this paper. Although
the rising phase could be very sharp, the decaying slope is rather
flat. This is because of two effects. First, after entering the
dense cloud, the blastwave Lorentz factor falls off rapidly. The
observed time $t \sim R/2 \Gamma^2$ is hence significantly
stretched. The dashed line in Fig.\ref{Fig:Cloud} represents the observed
emission before the fireball exits the cloud. Second, after the
fireball exits the cloud (dotted line in \ref{Fig:Cloud}), the
fireball does not decelerate immediately since the Lorentz factor is
already too low to be further decelerated in a medium with a low
density. It is decelerated again when enough medium is swept up at a
larger radius. The interaction of a fireball with density bumps has
been studied previously by many authors (e.g. Lazzati et al. 2002; Dai
\& Wu 2003). Our detailed calculations suggest that the variation
caused by density inhomogeneities is generally not very significant
(see also Ramirez-Ruiz et al. 2005). The lightcurves for a blastwave
surfing on a density wave should be generally quite smooth
(cf. Lazzati et al. 2002). 
The one-dimensional model presented here effectively calculates
the feature when the fireball encounters a high density shell. A more 
realistic model for density clumps should include the size of the 
clump ($\Delta$). The lightcurve decay slope after the peak therefore 
depends on the comparison between $\Gamma^{-1}$ and the angle the 
clump extends from the central engine, $\Delta / R$. If $\Gamma^{-1}
\ll \Delta/R$, the above calculation is still valid since the observer
would not notice the edge of the clump. The resulting decay slope is
rather shallow after the peak, which is in distinct contrast to the
X-ray flare data. This is the case for the
parameters we use in Fig. \ref{Fig:Cloud}. For smaller clumps,
the lightcurve will steepen as $\Gamma^{-1}$ becomes comparable to
$\Delta /R$. Our numerical calculations indeed show such a feature.
However, the flux does not fall rapidly right after the peak. The peak
time is typically shortly after the blastwave enters the density clump
when the Lorentz factor is still very high. The epoch when
$\Gamma^{-1}=\Delta /R$ is satisfied happens much later. The
lightcurve therefore still show a shallow decay 
segment before the steep decay component. This is in contrast with the
X-ray flare data that show rapid decays right after the flare peak
(e.g. Burrows et al. 2005; Falcone et al. 2005; Romano et al. 2005;
O'Brien et al. 2005). One could make the clouds small enough, so that 
$\Gamma^{-1} > \Delta/R$ is satisfied from the very
beginning. However, in such a case it is very hard to achieve a large
contrast between the flares and the underlying afterglow level (Ioka
et al. 2005). The 500 time contrast observed in GRB 050502B is in any
case very difficult to achieve within all the density clump models we
have explored. We therefore tentatively conclude that the X-ray flares
commonly detected in the early X-ray afterglow lightcurves are likely
not caused by the putative density clouds surrounding the GRB
progenitors. The lack of these well-modeled features in the data also
suggest that the lumpiness of circumburst medium is, if any, rather
mild. We note, however, Dermer (2005, in preparation) suggests that
more detailed 3-D modeling of the density clump problem could
potentially reproduce the observed X-ray flares.

\subsection{Two-component jet?}
\label{subsec:model4}

\begin{figure}
\epsscale{1.0}
\plotone{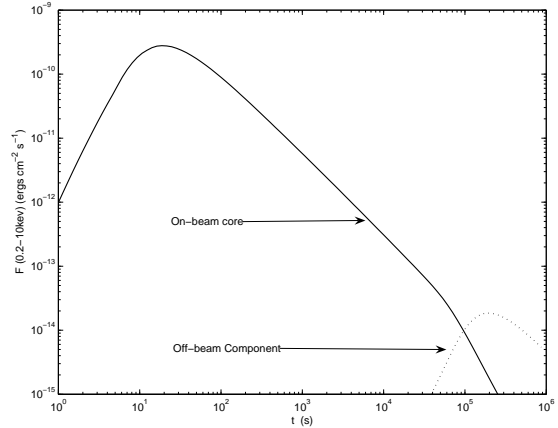}
\caption{The X-ray lightcurve powered by a two-component jet. The
angular range $0<\theta<0.1$ is the central narrow component, which
has $E_{\rm iso,n}=10^{52}$ergs and $\Gamma_{0,n}=240$. The wide jet
component covers a range of $0.1<\theta<0.3$, with $E_{\rm iso,w} =
5\times 10^{52}$ergs and $\Gamma_{0,w}=50$. The ISM density is 
$n=1{\rm cm^{-3}}$. Other parameters: $\epsilon_{\rm e}=0.1$,
$\epsilon_{\rm B}=0.01$, $p=2.3$, and $z=1$. The line of sight is at
$\theta=0$. The lightcurve peak of the second component corresponds to
the epoch when the off-beam wide component is decelerated, so that its
$1/\Gamma_{w}$ beam enters the field of view.}
\label{Fig:TwoComp}
\end{figure}

In the collapsar progenitor models, it is expected that the ejecta
generally has two components, one ultra-relativistic component
powering the GRB, and another moderately relativistic cocoon component
(e.g. Zhang. et al. 2004b; M\'esz\'aros \& Rees 2001; Ramirez-Ruiz et
al. 2002). This gives a physical motivation to the phenomenological
two-component jet model (e.g. Lipunov et al. 2001; 
Berger et al. 2003; Huang et al. 2004). An interesting possibility
would be whether the X-ray flare following the prompt gamma-ray
emission (which arises from the central relativistic component) is
caused by the deceleration of the wider mildly-relativistic cocoon
component as it interacts with the ambient medium. A straightforward
conclusion is that the wide, off-beam component must contain more
energy than the narrow component in order to give noticeable features
in the lightcurve. Also the decay after the lightcurve peak of the
second component should follow the standard afterglow model, and
the variability time scale satisfies $\delta t/t \sim 1$. The optical
lightcurves of the two component jet model have been calculated by
Huang et al. (2004) and Granot (2005). Fig.\ref{Fig:TwoComp} shows a
sample X-ray lightcurve in the two-component jet model. It is obvious
that this model cannot interpret the rapid fall-off observed in the
XRT X-ray flares, and is therefore ruled out.

\subsection{Patchy jets?}
\label{subsec:model5}

\begin{figure}
\epsscale{1.0}
\plotone{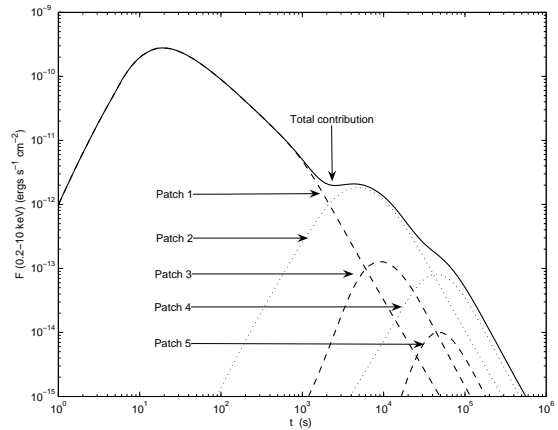}
\caption{The X-ray lightcurve powered by a patchy jet. For simplicity,
an annular patchy jet is simulated. Following parameters are adopted. 
Patch 1 (the on-beam jet): $0<\theta<0.02$, $E_{\rm iso,1}=10^{52}$ergs,
$\Gamma_{0,1}=240$; Patch 2: $0.02<\theta<0.04$, $E_{\rm iso,2}=5
\times 10^{52}$ergs, $\Gamma_{0,2}=50$; Patch 3: $0.04<\theta<0.06$,
$E_{\rm iso,3}=10^{52}$ergs, $\Gamma_{0,3}=240$; Patch 4:
$0.06<\theta<0.08$, $E_{\rm iso,4}=5 \times 10^{52}$ergs,
$\Gamma_0=50$; Patch 5: $0.08<\theta<0.10$, $E_{\rm
iso,5}=10^{52}$ergs and $\Gamma_{0,5}=240$. Other parameters are the
same as Fig.\ref{Fig:TwoComp}.}
\label{Fig:Patchy}
\end{figure}

A related model considers a jet with large energy fluctuations in
the angular direction, so that its energy distribution is patchy
(Kumar \& Piran 
2000b). This is a variation of the two-component jet, and could be
approximated as a multi-component jet. When the $1/\Gamma$ cone of
different patches enter the field of view, the observed lightcurve may
present interesting signatures. However, the general feature of the
two-component jet still applies: Only when a patch has a substantial
energy compared with the on-beam jet would it give a bump feature on
the lightcurve. After each bump, the afterglow level is boosted and
would not resume the previous level. The variability time scale is
also typically $\delta t/t \sim 1$. Figure \ref{Fig:Patchy} gives an
example. Apparently this model can not account for the observed X-ray
flares. 
\subsection{Post energy injection into the blastwave?}
\label{subsec:model6}

In the internal shock model, it is expected that after the collisions are
finished, the shells are distributed such that a shell with a higher
Lorentz factor always leads a shell with a lower Lorentz factor. As the
fast moving shell (blastwave) is decelerated, the trailing slow shell
will catch up with it and inject energy into the blastwave (Kumar \&
Piran 2000a). Such an injection also happens if the central engine
further ejects high-$\Gamma$ shells that catch up with the
decelerating blastwave (Zhang \& M\'esz\'aros 2002b). Such a collision
would give rise to a bump signature on the lightcurve. A detailed
treatment suggests that the overall lightcurve should include emission
from three components: a forward shock propagating into the medium, a
second forward shock propagating into the blastwave, and a reverse
shock propagating into the injected shell. In the X-ray band, the
contribution from the reverse shock is negligible. The lightcurve
generally shows a step-like signature, due to the increase of the
total energy in the blastwave (Zhang \& M\'esz\'aros 2002b). 
After the peak, the flux level does not resume the previous level
since more energy has been injected into the fireball. The decay slope
after the injection peak follows the standard afterglow model, and
$\delta t/t \sim 1$ is expected. These are also inconsistent with the
data of X-ray flares.

\subsection{Neutron signature?}

Another interesting possibility is whether X-ray flashes are the
signature of the existence of free neutrons in the fireball. 
Derishev et al. (1999), Beloborodov (2003) and Rossi et al. (2004)
suggested that a baryonic fireball contains free neutrons, the decay
of which would leave important imprints on the lightcurve. Fan et
al. (2005a) modeled the process carefully and calculated the
lightcurves. According to Fan et al. (2005), the neutron-feature is
rather smooth in a wind model, and it is hard to detect. In the ISM
case, on the other hand, a bump does exist (in all bands). The
physical reason is that the trailing proton shell catches up with the
decelerated neutron-decay products. The physical process is analogous
to the post energy injection effect discussed in \S\ref{subsec:model6}. 
The amplitude of the flare is modest, at most a factor of
several. Since the injection model is not favored, this possibility is
also disfavored.

\subsection{Late central engine activity}
\label{subsection:Model7}

After ruling out various ``external-origin'' mechanisms, we are left
only with the possibility that involves the re-activation of the
central engine. In this interpretation, the X-ray flares share
essentially the same origin as the prompt gamma-ray emission, i.e. they
are caused by some ``internal'' energy dissipation processes which
occur before
the ejecta is decelerated by the ambient medium. The leading scenario
is the ``late'' internal shock model, which suggests that the central
engine ejects more energy in the form of an unsteady wind with varying
Lorentz factors at a late time. These discrete shells collide with each
other and produce the observed emission. Alternatively, the late
injection could be mainly in the form of magnetic fields/Poynting
flux, and the X-ray flares are due to the intermittent dissipation of
the magnetic fields, likely through magnetic reconnection events.
Fan, Zhang \& Proga (2005c) argued that at least for the flares
following short GRBs, the process that powers the flares has to be
magnetic-origin. 

There are two advantages of the ``internal'' models over the
``external'' models. 

First, re-starting the central engine equivalently re-set the time 
zero point. In this interpretation, the observed 
flare component and the underlying decaying component observed at the
same observer time $t$ originate from
different physical sites at different central engine times. Let us
assume that the initial burst lasts $T_{90}$, that the central engine
re-activates after a time interval $\Delta t$, and that it ejects an
unsteady wind with a typical variability time scale of $\delta
t$. At the observer time $t=T_{90}+\Delta t \sim 1000$ s, the
underlying decaying component (external afterglow) happens at a
distance $R_{ex} \sim [\Gamma(t)]^2 c t \sim 5\times 10^{16}$ cm,
where $\Gamma(t)$ is the Lorentz factor of the blastwave at the time
$t$. The flare, on the other hand, happens at a distance of $R_{in}
\sim \Gamma_0^2 c \delta t \sim (10^{13}-10^{14})$ cm, where
$\Gamma_0$ is the initial Lorentz factor of the late-time
ejecta. According to the clocks attached to 
the central engine, the photons from the external afterglow component
are emitted at $\hat t_{ex} \sim R_{ex} / c \sim 1.7\times 10^6$s,
while the photons from late central engine activity are emitted at
$\hat t_{in} \sim t+R_{in}/c  
\sim 10^3 + R_{in}/c \sim (10^3 - 10^4)$s. Because of the relativistic
effect, these photons reach the observer at exactly the same time $t$,
and superpose onto the lightcurve detected by the observer. When
plotted as a single $\log-\log$ lightcurve with the origin at the burst
trigger, a very steep apparent decay slope can be produced for a large
time shift $\Delta t$ (Fig.\ref{Fig:Internal}). This naturally
overcomes the $\delta t/t \geq 1$ constraint encountered by the
external models (Ioka et al. 2005 and references therein).

Second, invoking a late central engine activity greatly eases the
required energy budget. In most of the external models, in order to
give rise to a significant bump on the lightcurve, the total
newly-added energy (either from the radial direction - late injection
case, or from the horizontal direction - patchy jets or
multiple-component jets) must be at least comparable to the energy
that defines the afterglow emission level. This model therefore
demands a very large energy budget. For the internal model, on the
other hand, since the lightcurve is a superposition of two independent
physical components, the energy budget is greatly reduced, especially
if the bump happens at later times when the background afterglow level
is much lower. For example, for an X-ray lightcurve with the decay
index of -1 following a burst with duration 10 s, a significant flare
at $\sim 10^4$s only requires a luminosity slightly larger than
$10^{-3}$ times of that of the prompt emission. 
This model is therefore very ``economical'' in interpreting the very
late ($>10^4$s) flares detected in some bursts (e.g. GRB 050502B,
Falcone et al. 2005; and GRB 050724, Barthelmy et al. 2005b).

Can late central engine activity give rise to softer bursts
(e.g. X-ray flares as compared with the prompt gamma-ray or hard X-ray
emission)? According to the internal shock model, the peak energy
$E_p$ of the synchrotron spectrum satisfies (Zhang \&
M\'esz\'aros 2002c)
\be
E_p \propto \Gamma B' \propto L^{1/2} r^{-1} \propto L^{1/2}
\Gamma^{-2} \delta t^{-1}~, 
\ee
where $L$ is the luminosity, $B'\propto (L/\Gamma^2r^2)^{1/2}$ is the
comoving magnetic field strength, $\Gamma$ is the typical Lorentz
factor of the wind (for internal shock collisions, this $\Gamma$ is
for slow shells), and $\delta t$ is the variability time scale. We can
see that a smaller $L$, a higher $\Gamma$ and a larger $\delta t$ would be
favorable for a softer burst. The observed XRT X-ray flares generally
have a smaller luminosity $L$. The lightcurve is smoother with a
larger $\delta t$ (Burrows et al. 2005; Barthelmy et al. 2005b). Also,
at later times, the environment tends to be cleaner so that $\Gamma$
could be larger. One therefore naturally expects softer flares at
later times. 

Can the central engine re-start after being quiescent for some time
(e.g. $\sim 10^3$s, but sometimes as late as $\sim 10^5$s)? This is
an interesting theoretical problem. The collapsar model predicts a
central engine time scale of minutes to hours (MacFadyen et
al. 2001). The star may also fragment into several pieces during the
collapse (King et al. 2005). The largest piece collapses first onto
the black hole, powering a prompt gamma-ray burst. Other fragments
are intially ejected into elliptical orbits, but would eventually fall
into the black hole after some time, powering the late X-ray
flares. Fragmentation could also happen within the accretion disk
itself due to gravitational instabilities (Perna et al. 2005). A
magnetic-dominated accretion flow could also give rise to intermittent
accretion flows due to interplay between the gravity and the magnetic
barrier (Proga \& Begelman 2003; Proga \& Zhang 2005). Flares could
also occur in other central engine scenarios (Dai et al. 2005). The rich
information collected by the {\em Swift} XRT suggests that we are
getting closer to unraveling of the details of the bursting mechanisms
of GRBs. More detailed studies are called for to unveil the mystery of
these explosions.

An interesting fact (see next subsection for case studies) is that the
duration of the flares are positively correlated with the time at
which the flare occurs. The later the flare, the longer the flare
duration. A successful central engine model must be able to address
such a peculiar behavior. Perna et al. (2005) suggest that if the
accretion disk is fragmented into blobs or otherwise has large
amplitude density fluctuation at large radii from the central engine,
the viscous disk evolution would cause more spread for blobs further
out from the central engine. This gives a natural mechanism for the
observed correlation. Perna et al. (2005) suggest that gravitational
instability in the outer part of the disk is likely the origin of the
density inhomogeneity within the disk.

\subsection{Case studies}

In this subsection, we briefly discuss the X-ray flares discovered in
several GRBs.

{\bf GRB 050406 \& GRB 050502B} (Burrows et al. 2005): These were the
first two bursts with flares detected by XRT. For the case of GRB
050406 whose 
$T_{90}$ is $\sim 5$s, an X-ray flare starts at $\sim 150$s and
reaches a peak at $\sim 230$s. The 
flux rebrightening at the flare peak is by a factor of 6. The rising
and the decaying indices are $\sim 4.9$ and $-5.7$, respectively. The
total energy emitted during the flare is about 10\% of that emitted in
the prompt emission. This suggests that the central engine becomes
active again after $\sim 150$s, but with a reduced power. For the case
of GRB 050502B whose $T_{90}$ is $\sim 17.5$s, a giant flare starts at
$\sim 300$s and reaches a peak at $\sim 740$s. The flux rebrightening
is by a factor of $\sim 500$. The decay index after the peak is
$\sim -6$. The total energy emitted during the
flare is comparable to that emitted in the prompt emission. This
suggests that the central engine re-starts and ejects a substantial
amount of energy. The duration of the X-ray flare is much longer
that the duration of the prompt emission ($T_{90}$), so that the
luminosity of the flare is much lower. In GRB 050502B, there is yet
another late flare-like event that peaks at $\sim 7\times 10^4$s. The
post-peak decay index is $\sim -3$. This is also consistent with the
case of a late-time central engine activity, and the decaying slope is
consistent with that expected from the curvature effect. The total
energy emitted in this flare is $\sim 10\%$ of that emitted in the
giant flare at $\sim 740$s.

{\bf GRB 050724} (Barthelmy et al. 2005b): This is a short, hard burst
with $T_{90} \sim 3$s whose host galaxy is an elliptical galaxy,
similar to the case of the first {\em Swift}-localized short burst
GRB 050509B (Gehrels et al. 2005). The XRT lightcurve reveals rich
features which are quite similar to the case of GRB 050502B. 
The XRT observation starts at $\sim 74$s after the trigger, and the
early XRT lightcurve initially shows a steep decay with a slope 
$\sim -2$. This component is connected to the extrapolated BAT
lightcurve that shows a flare-like event around $(60-80)$s. This
extended flare-like epoch (including the -2 
decay component) stops at $\sim 200$s after which the lightcurve
decays even more rapidly (with an index $<-7$). A second, less
energetic flare peaks at $\sim 300$s, which is followed by
another steep (with index $<-7$) decay. A third, significant
flare starts at $\sim 2\times 10^4$s, and the decay index after the
peak is $\sim -2.8$. This quite similar to those the late flare 
seen in GRB 050502B. The energy emitted during the third flare is
comparable to that of the second one, and both are several times
smaller than the energy of the first flare. All the rapid-decay
components following the flares could be potentially interpreted as
the high-latitude emission, 
given a proper shift of the zero time point. Since the duration of the
third flare is very long, the $t_0$ effect does not affect the decay
index too much. This is why the decay index $-2.8$ is quite
``normal'', i.e. consistent with the $-2-\beta$ prediction. The
earlier steeper decays (e.g. $<-7$) are all preceded by flares with
a sharp increasing phase. Shifting $t_0$ in these cases would lead to
significant flattening of the decaying index, which could be still
consistent with the high-latitude emission. All these discussions also
apply to the case of GRB 050502B\footnote{After submitting this paper,
detailed data analyses (Liang et al. 2005) confirm these speculations.
The late flares in both GRBs 050724 and 050502B are consistent
with the hypothesis that they are due to late time central engine
activity.}.

{\bf GRB 011121} (Piro et al. 2005). This {\em BeppoSAX} burst also
indicated a flare-like event around $\sim 270$s. Piro et al. (2005)
interpreted the X-ray bump as the onset of the afterglow phase. In
view of the fact that X-ray flares are commonly detected in {\em
Swift} bursts, it is natural to spectulate that the event is also an
X-ray flare caused by late central engine activity. Fan \& Wei (2005)
have suggested the late central engine activity and performed a
detailed case study on this event. 


\section{Conclusions}
\label{sec:conclusions}

During the past several months, the {\em Swift} XRT has collected
a rich sample of early X-ray afterglow data. This, for the first time,
allows us to peer at the final temporal gap left by the previous
observations and to 
explore many interesting questions of GRB physics. In this paper,
we have systematically investigated various possible physical processes
that could give interesting contributions to the early X-ray afterglow
observations. This includes the tail emission of the prompt gamma-ray
emission, both the forward shock and the reverse shock emission
components, refreshed shocks, post energy injection, medium
density clumps near the burst, angular inhomogeneities of the
fireball, emission component due to the presence of free neutrons, as
well as emission from late central engine activity. We discuss how
the above processes might leave interesting signatures on the early
X-ray afterglow lightcurves.

Based on the XRT data collected so far, we summarize the
salient features and suggest a
tentative synthetic lightcurve for the X-ray afterglow
lightcurves. As shown in Fig. \ref{XRTlc}, the synthetic lightcurve
includes five components: I. an intial steep decay component;
II. a shallow-than-normal decay component; III. a ``normal'' decay
component; IV. a post-jet break component; and V. X-ray flares. 
The components I and III appear in almost all the 
bursts. Other three components also commonly appear in some bursts.
Flares have been detected in nearly half of the XRT early
lightcurves, and the shallow decay segment has also been discovered
in a good fraction of {\em Swift} GRBs. We therefore believe that they
represent some common underlying physics for GRBs. After comparing
data with various physical models, we tentatively draw the
following conclusions.

1. The rapid decay component (Tagliaferri et al. 2005) commonly
observed in the very early afterglow phase (which usually has a
different spectral slope than the late shallow decay components) is
very likely the tail 
emission of the prompt gamma-ray bursts or of the early X-ray flares.
Allowing proper shifting of the time zero point and considering the
contribution of the underlying forward shock emission, we speculate
that essentially all the steep decay cases could be understood in
terms of the ``curvature effect'' of the high-latitude emission as the
emission ceases abruptly. More detailed data analyses (Liang et
al. 2005) support such a speculation.

2. The transition between the prompt emission and the afterglow
emission appears to be universally represented by a rapid decay
followed by a shallower decay, indicating that the GRB emission site
is very likely different from the afterglow site, and that the
apparent gamma-ray efficiency is very high.

3. In a good fraction of GRBs (e.g. GRBs 050128, 050315, 050319, 050401,
Campana et al. 2005; Vaughan et al. 2005; Cusumano et al. 2005; de
Pasquale et al. 2005), a clear temporal break exists in the early
X-ray lightcurves. There is 
no obvious spectral index change across the break. The temporal decay
index before the break is very flat, while that after the break is
quite ``normal'', i.e. is consistent with the standard afterglow
model for a fireball with constant energy expanding into an ISM. We
suggest that these breaks are likely not ``jet breaks''. Rather they
mark the cessation of an early continuous energy injection phase
during which the external shock is refreshed. We suggest three
possible physical mechanisms for the refreshed shocks, i.e. a
long-lived central engine with a decaying luminosity, a power law
distribution of the shell Lorentz factors before deceleration begins,
and the deceleration of a high-$\sigma$ flow. Further studies are
needed to better understand this phase.

4. In the ``normal'' phase, the data for many bursts are consistent
with an ISM medium rather than a wind medium. This has important
implications for understanding the massive star progenitors of long
GRBs, including their late evolution stage shortly before explosion.

5. Given that most of the shallow-to-normal transitions are due to the
cessation of the refreshed shock phase, the cases with a
well-identified jet break are not very common. Nonetheless, jet breaks
are likely identified in some bursts, e.g. GRB 050315 (Vaughan et
al. 2005), GRB 050525A (Blustin et al. 2005), XRF 050416 
(Sakamoto et al. 2005), and some others.

6. The X-ray flares detected in nearly half of the {\em Swift} bursts
are most likely due to late central engine activity, which results
in internal shocks (or similar energy dissipation events) at later
times. It seems to us that there is no evidence for the existence of
density clumps in the GRB neighborhood, and there is no support for
strong angular inhomogeneities (e.g. two-component jet, patchy jets)
for the GRB fireball. However, their existence is not ruled out.

7. The similar lightcurves for some long GRBs (e.g. GRB050502B,
Burrows et al. 2005) and some short GRBs (e.g. GRB050724, Barthelmy et
al. 2005) indicate that different progenitor systems may share some
similarities in the central engine. This might be caused by the same
underlying physics that controls the hyper-accreting accretion disk -
a common agent in charge both types of systems (Perna et al. 2005).

The {\em Swift} XRT is still rapidly accumulating data on the early
X-ray afterglows. More careful statistical analyses of these X-ray data,
as well as more detailed case studies, also including low frequency
data collected by UVOT and other ground-based telescopes, would greatly
imporve our knowledge about GRB prompt emission and afterglows,
advancing the quest for the final answers to the core questions in the
study of GRBs.


\acknowledgments 
We thank the referee for helpful comments.
B.Z. acknowledges useful discussions with T. Abel, P. Armitage,
M. Begelman, Z. G. Dai, C. D. Dermer, B. Dingus, C. Fryer, L. J. Gou,
J. Granot, A. Heger, D. Lazzati, A. Panaitescu, R. Perna, D. Proga,
S. Woosley, X. Y. Wang, X. F. Wu, and W. Zhang on various topics
covered in this paper. 
This work is supported by NASA NNG05GB67G, NNG05GH91G (for BZ),
NNG05GH92G (for BZ, SK and PM), Eberly Research Funds of Penn State
and by the Center for Gravitational Wave Physics under grants
PHY-01-14375 (for SK), NSF AST 0307376 and NASA NAG5-13286 (for PM),
and NASA NAS5-00136 (DB \& JN).

\end{document}